\documentclass[preprint,titlepage,prx,amsmath,amssymb,aps,floatfix, nofootinbib]{revtex4-2}
\usepackage{graphicx} 
\usepackage{amsmath}
\usepackage{braket}
\usepackage{hyperref}
\usepackage{cleveref}
\usepackage{xcolor}
\usepackage{textcomp}
\usepackage{lmodern}
\usepackage[T1]{fontenc}
\usepackage{gensymb}
\newtheorem{claim}{Claim}
\newtheorem{proof}{Proof}

\begin{document}

\title{Type-I and Type-II Fusion Protocols for Weighted Graph States}
\author{Noam Rimock}
\email[]{noamrimock@gmail.com}
\author{Yaron Oz}
\email[]{yaronoz@tauex.tau.ac.il}
\affiliation{School of Physics and Astronomy, Tel Aviv University, Ramat Aviv 69978, Israel}

\date{\today}

\begin{abstract}
Weighted graph states extend standard graph states by associating phases with entangling edges, and may serve as resources for measurement-based quantum computation (MBQC). 
We analyze how the two main fusion operations, Type-I and Type-II, act on weighted graph states. 
Type-I fusion operates identically to the unweighted case, merging two one-dimensional weighted graphs, while preserving edge weights and success probabilities. In addition, the pool of 2-qubit weighted graph states can be generated easily from GHZ states or Bell pairs.
In contrast, Type-II fusion requires a logical qubit, which can be formed only for specific weight configurations, and with a success probability below one-half.
When successful, it fuses the states correctly, but its failure outcomes destroy the structure of the graphs, removing the good-failure feature known from ordinary graph states. 
These two obstacles can be overcome by using extra qubits and $\pi$-weights in the 1-dim weighted graph states that are being fused, which enables scalability.
We compute the change in the entanglement entropy of the resulting link due to the fused states being weighted graph states (for generalized fusion), and classify the resulting states of a general non-Bell projection.
\if{We present a general rule for generating target weighted edges via measurement, and quantify the entanglement reduction due to non-orthogonality. }\fi
These results define the practical limits of the fusion-based construction of weighted graph states. 
To our knowledge, this is the first time that fusion processes have been used to construct weighted graph states.
\end{abstract}

\maketitle

\section{Introduction}

Graph states constitute a central class of entangled quantum states that form the basis of measurement-based quantum computation (MBQC) \cite{Briegel_2001}. 
In this framework, computation is performed through single-qubit measurements on a highly entangled resource, typically a cluster state. 
The efficient generation and manipulation of such resource states are therefore key ingredients for scalable quantum computing, particularly in photonic architectures, where entanglement is created probabilistically. 

A promising approach to constructing large graph states is the use of fusion operations, probabilistic entangling measurements that connect smaller cluster fragments. 
Two main types of fusion gates, Type-I and Type-II, have been introduced for standard (unweighted) graph states \cite{Browne_2005}. 
Type-I fusion probabilistically merges two graph states by removing the measured qubits, while preserving the connectivity of their neighbors, whereas Type-II fusion enables the creation of logical qubits, and supports fault-tolerant graph growth through its favorable failure behavior. 

Weighted graph states \cite{hein2006entanglementgraphstatesapplications} generalize standard graph states by assigning continuous phase factors to their edges. 
These states appear naturally in various physical implementations and areas of research \cite{Hartmann_2007,Anders_2007,PhysRevA.78.042332,7zxj-jp34}, extend the resource space available for MBQC \cite{Gross_2007,yamazaki2025measurementbasedquantumcomputationweighted}, and have already been realized experimentally \cite{skosana2026experimentalrealizationphotonicweighted}.  
However, the behavior of fusion operations in the weighted setting has not yet been discussed, let alone systematically analyzed. Today, the only theoretical way to construct an optical weighted graph state is by applying the appropriate 2-qubit gates. To date, the only experimentally realized optical weighted graph states are consist of two qubits \cite{skosana2026experimentalrealizationphotonicweighted}.

In this work, we study how Type-I and Type-II fusions translate to the case of weighted graph states. 
We show that Type-I fusion acts identically to the unweighted case, maintaining both the success probability and edge-weight inheritance. 
In contrast, Type-II fusion exhibits significant modifications: the creation of a logical qubit becomes conditional on specific weight configurations, and its non-successful outcomes do not preserve the desired graph structure. Those limitations can be bypassed at a cost of $O(n\ln{n})$ additional qubits as explained later, where $n$ is the number of needed successful Type-II fusion processes. 
We also derive conditions under which measurements on GHZ-states can deterministically generate 2-qubit weighted graph states with a target weight that does not depend on the outcome of the measurement, hence creating a pool of 2-qubit states that one needs in order to create larger weighted graph states. Finally, we study the behavior of weighted graph states under generalized fusion, classify the resulting states, {\color{black}and compute the new probabilities and entanglement entropies that change due to non-maximal edge weights. We prove that any Type-II fusion process of graph states involving projection onto maximally entangled states will retain its success probability when applied to weighted graph states instead.}

Together, these results clarify the limitations and applicability of fusion-based schemes for constructing weighted graph states, and fusion gates can now be applied to experimentally realize weighted graph states with more than two qubits.

The remainder of this paper is organized as follows. 
Section~\ref{Section:Reviewing Known Theory} reviews the theoretical background, including the definitions of optical qubits, weighted graph states, and the standard Type-I and Type-II fusion protocols, as well as the construction of logical qubits and the generalized fusion framework. 
Section~\ref{Section:Fusions for weighted graph state} presents our main results, analyzing how fusion operations behave in weighted graph states and detailing the conditions for successful logical-qubit creation and entanglement generation, as well as studying the result of a non-Bell projection and generalized fusion on weighted graph states. 
In Section~\ref{Section:Discussion}, we summarize our findings, discuss their implications for measurement-based quantum computation, and outline possible directions for future work.

\section{Preliminaries: Weighted Graph States and Fusion Protocols}
\label{Section:Reviewing Known Theory}
In this section, we review the theoretical background necessary for the analysis presented in the following sections. 
We begin by defining optical qubits, and recalling the construction of weighted graph states, which generalize standard graph states by introducing phase-dependent entangling edges. 
We then summarize the known fusion mechanisms, Type-I and Type-II, that probabilistically connect graph-state fragments, and explain their realization in linear-optical systems. 
Finally, we describe how logical qubits can be created within one-dimensional graph states, and outline the generalized Type-II fusion and non-Bell projection schemes, that extend these standard fusion protocols. 
This overview establishes the notation and mathematical tools used throughout the remainder of the paper.


\subsection{Optical qubits}
\label{subsec:Optical qubit}
Given a photonic frequency/position channel $a$, with two polarizations $H$, $V$, we use it to represent a qubit $\ket{\phi}=\alpha\ket{0}_a+\beta\ket{1}_a$, where:

\begin{gather}
    \ket{0}_a=a_H^{\dagger}\ket{vac},\hspace{0.1cm}\ket{1}_a=a_V^{\dagger}\ket{vac} \ .
    \label{eq:RepresentingZeroAndOneByHorizontalAndVertical}
\end{gather}
That way, the mathematical representation of linear optical devices that do not change the number of photons is convenient. 

\subsection{Weighted graph states}
\label{subsec:Weighted graph states}
Weighted graph states are quantum states associated with weighted graphs \cite{hein2006entanglementgraphstatesapplications,Hartmann_2007}. Given a weighted graph, a qubit is assigned in the initial state $\ket{+}=\frac{1}{\sqrt{2}}\left(\ket{0}+\ket{1}\right)$ for every vertex in the graph, and then the operator $e^{-i\chi_{ab}\ket{1}_a\ket{1}_b\bra{1}_a\bra{1}_b}$ is applied on every pair of qubits $a,b$ whose associated vertices are connected by an edge with weight $\chi_{ab}$, yielding the state:

\begin{gather}
    \label{eq:DefinitionOfWeightedGraphState}
    \ket{\phi}=\prod_{(a,b)\in E} e^{-i\chi_{ab}\ket{1}_a\ket{1}_b\bra{1}_a\bra{1}_b} \prod_{c\in V} \ket{+}_c \ ,
\end{gather}
where $V$ is the set of the vertices of the graph and $E$ is the set of the edges of the graph. The order in which the $e^{-i\chi_{ab}\ket{1}_a\ket{1}_b\bra{1}_a\bra{1}_b}$ gates are applied is not relevant, as they commute with each other. There exists an equivalent recursive definition of weighted graph states: start with a single qubit in the state $\ket{+}$, and at each step we add a vertex and its edges to the existing vertices. At every step, if the wave function of the current weighted graph state is $\ket{\phi}$ and we add the qubit $a$ with weights $\chi_{ab}$ on the edges to all its existing neighbors $b\in n(a)$, then the new wave function is given by:

\begin{gather}
    \label{eq:RecursiveDefinitionOfWeightedGraphState}
    \ket{\psi}=\frac{1}{\sqrt{2}}\left(\ket{0}_{a}\ket{\phi}+\ket{1}_a\prod_{b\in n(a)}e^{-i\chi_{ab}\ket{1}_b\bra{1}_b}\ket{\phi}\right) \ .
\end{gather}
In the case where all the weights are $\chi_{ab}=\pi$, the state (\ref{eq:DefinitionOfWeightedGraphState}) is called a graph state (which is associated with the given graph):

\begin{gather}
    \label{eq:DefinitionOfGraphState}
    \ket{\phi}=\prod_{(a,b)\in E} CZ^{a,b} \prod_{c\in V} \ket{+}_c \ ,
\end{gather}
where $CZ^{a,b}$ is the controlled-Z gate acting on the qubits $a$ and $b$. The recursive equation (\ref{eq:RecursiveDefinitionOfWeightedGraphState}) becomes:

\begin{gather}
    \label{eq:RecursiveDefinitionOfGraphState}
    \ket{\psi}=\frac{1}{\sqrt{2}}\left(\ket{0}_{a}\ket{\phi}+\ket{1}_a\prod_{b\in n(a)}Z_{b}\ket{\phi}\right) \ .
\end{gather}

Graph states are important because MBQC can be realized on 2-dim graph states \cite{Briegel_2001}. There also exist examples of two-dimensional weighted graph states that serve as suitable resource states for MBQC \cite{Gross_2007,yamazaki2025measurementbasedquantumcomputationweighted}; however, the full set of weighted graph states that are universal (and can be used for MBQC) is still unknown.

Given a graph, weighted graph states are all states obtained by acting on a set of qubits (one for each vertex in the graph) with two-qubit commuting gates on every pair of qubits connected by an edge, up to single-qubit rotations \cite{hein2006entanglementgraphstatesapplications}.

\subsection{Type-I Fusion}
\label{Subsec:Fusion Type-I}
Given two one-dimensional graph states and qubits $a,b$ at the ends of those graph states, Type-I fusion is a probabilistic gate that, when successful, erases both qubits $a,b$, and a new qubit $c$ is created, which is connected to the previous neighbors of $a$ and $b$. This results in a single one-dimensional graph state whose length is the sum of the lengths of the original graph states, minus one qubit. When the fusion fails, the qubits $a,b$ are still erased, but the graph states remain separated. In the following paragraphs, we briefly explain the mathematics of this process.

By equation (\ref{eq:RecursiveDefinitionOfGraphState}), the wave function of the two original graph states can be written as (up to normalization):

\begin{gather}
    \label{eq:StartingWaveFunctionFusionTypeIGraphStates}
    \ket{\psi}=\left(\ket{0}_{a}\ket{\phi_a}+\ket{1}_a\prod_{e\in n(a)}Z_{e}\ket{\phi_a}\right)\left(\ket{0}_{b}\ket{\phi_b}+\ket{1}_b\prod_{f\in n(b)}Z_{f}\ket{\phi_b}\right) \ .
\end{gather}
When the fusion succeeds (with probability one-half, which can be increased using ancilla qubits), one of the following operators is applied on the state (\ref{eq:StartingWaveFunctionFusionTypeIGraphStates}) (up to normalization):

\begin{gather}
    \label{eq:OperatorFusionTypeISuccess}
    \ket{0}_c\bra{0}_a\bra{0}_b\pm \ket{1}_c\bra{1}_a\bra{1}_b \ ,
\end{gather}
yielding the state:

\begin{gather}
\label{eq:FusedGraphStateInCaseOfSuccessOfFusionTypeIForGraphState}
    \ket{\psi_{new}}=\frac{1}{\sqrt{2}}\left(\ket{0}_{c}\ket{\phi_a}\ket{\phi_b}\pm \ket{1}_c\prod_{e\in n(a)}Z_{e}\prod_{f\in n(b)}Z_{f}\ket{\phi_a}\ket{\phi_b}\right) \ .
\end{gather}
If the sign is negative, one can apply a $Z_c$ operator to change the sign to positive, so in both cases we end up with the plus sign, which by equation (\ref{eq:RecursiveDefinitionOfGraphState}) is the desired fused graph state.

In the case of failure (which has probability one-half, which can be lowered by using ancilla qubits), one of the following operators is applied on the state (\ref{eq:StartingWaveFunctionFusionTypeIGraphStates}) (up to normalization):

\begin{gather}
    \ket{0}_c\ket{1}_c\bra{0}_a\bra{0}_b \ , or \hspace{0.1cm} \bra{1}_a\bra{0}_b \ .
\end{gather}
We can ignore the optional creation of the non-computational state $\ket{0}_c\ket{1}_c$ that might multiply our wave function, and we get for both options "$Z$ measurements" on both qubits $a,b$, which results in their elimination, along with their connections, without affecting the other qubits in the two original graph states that remain separated.

To understand how Type-I fusion is performed and how to compute the probabilities of the different outcomes, we denote $\ket{\phi_a}$ by $f_1$, $\prod_{e\in n(a)}Z_{e}\ket{\phi_a}$ by $f_2$, $\ket{\phi_b}$ by $f_3$ and $\prod_{f\in n(b)}Z_{f}\ket{\phi_b}$ by $f_4$. 
We consider the creation operators $a_H^{\dagger},a_V^{\dagger},b_H^{\dagger},b_V^{\dagger}$ of the channels representing the qubits $a,b$ (as in equation (\ref{eq:RepresentingZeroAndOneByHorizontalAndVertical})). We can write the state (\ref{eq:StartingWaveFunctionFusionTypeIGraphStates}) as

\begin{gather}
\label{eq:StartingWaveFunctionFusionTypeIGraphStatesInf1f2f3f4Language}
    \ket{\psi}=\left(f_1a_H^{\dagger}+f_2a_V^{\dagger}\right)\left(f_3b_H^{\dagger}+f_4b_V^{\dagger}\right)\ket{vac} \ .
\end{gather}
Using only linear optical elements, any $4$ by $4$ unitary can be implemented on $Span\{{a_H^{\dagger},a_V^{\dagger},b_H^{\dagger},b_V^{\dagger}\}}$, yielding a new basis $c_H^{\dagger},c_V^{\dagger},d_H^{\dagger},d_V^{\dagger}$ \cite{CreatingU0}. Specifically, we transfer $a,b$ using a polarizing beam splitter, with one output being the $c$ channel, and the other output going through $45\degree$ rotation --- and the resulting channel being the $d$ channel. The state (\ref{eq:StartingWaveFunctionFusionTypeIGraphStatesInf1f2f3f4Language}) is now given in the representation \cite{Kok_2009}:

\begin{gather}
\label{eq:StartingWaveFunctionFusionTypeIGraphStatesInf1f2f3f4LanguageAfterTransferingToCandD}
    \ket{\psi}=\left(\frac{f_1f_3}{\sqrt{2}}c_H^{\dagger}\left(d_H^{\dagger}+d_V^{\dagger}\right)+f_1f_4c_H^{\dagger}c_V^{\dagger}+\frac{f_2f_3}{2}\left({d_H^{\dagger}}^2-{d_V^{\dagger}}^2\right)+\frac{f_2f_4}{\sqrt{2}}c_V^{\dagger}\left(d_H^{\dagger}-d_V^{\dagger}\right)\right)\ket{vac} \ .
\end{gather}

Now we measure the channels $d_H,d_V$. When computing, for instance, the probability to measure one qubit in $d_H$ channel and zero qubits in $d_V$ channel, one takes $\ket{\psi}$ in equation (\ref{eq:StartingWaveFunctionFusionTypeIGraphStatesInf1f2f3f4LanguageAfterTransferingToCandD}) and projects it onto $d_H^\dagger\ket{vac}\otimes H$, where $H$ is the Hilbert space consisting of all the other qubits from the two original graph states that are not $a,b$ (or even its subspace $Span\{f_1,f_2\}\otimes Span\{f_3,f_4\}$). The probability of this outcome is $\frac{1}{4}$, and the resulting state is (up to normalization) $\left(f_1 f_3c_H^{\dagger}+f_2 f_4c_V^{\dagger}\right)\ket{vac}$, which exactly operates with the plus option of eq. (\ref{eq:OperatorFusionTypeISuccess}) on the state (\ref{eq:StartingWaveFunctionFusionTypeIGraphStates}), getting the fused graph wave function (\ref{eq:FusedGraphStateInCaseOfSuccessOfFusionTypeIForGraphState}). 
{\color{black}Note that the computation of those probabilities doesn't assume that $f_1,f_2$ and $f_3,f_4$ are orthogonal couples, but only that they are in different Hilbert spaces than $Span\{{a_H^{\dagger},a_V^{\dagger},b_H^{\dagger},b_V^{\dagger}\}}$, and that the norms of $f_1,f_2$ are the same, as are the norms of $f_3,f_4$. This is because, when measuring one photon, we are left with a projected expression proportional to $\left(f_1 f_3c_H^{\dagger}\pm f_2 f_4c_V^{\dagger}\right)\ket{vac}$, whose norm is independent of whether $f_1,f_2$ or $f_3,f_4$ are orthogonal, because $c_H^\dag \ket{vac}, c_V^\dag \ket{vac}$ are orthogonal. When measuring two or zero photons, we are left with a projected expression proportional to $f_1 f_4 c_H^{\dag} c_V^{\dag} \ket{vac}$ or $f_2 f_3$, with norms that again do not depend on $f_1,f_2$ or $f_3,f_4$ being orthogonal.}


\if{Note that the computation of those probabilities doesn't assume anything about $f_1,f_2,f_3,f_4$ rather than the fact that they are in different Hilbert spaces than $Span\{{a_H^{\dagger},a_V^{\dagger},b_H^{\dagger},b_V^{\dagger}\}}$.}\fi

\subsection{Creating a logical qubit}
\label{Subsec:Creating a logical qubit}
To perform Type-II fusion, one needs to create a logical qubit in a one-dimensional graph state --- meaning that one of the qubits, $L$, is in fact two qubits entangled together, $b_1,b_2$, such that:

\begin{gather}
\label{eq:LogicalQubit}
    \ket{0}_L=\ket{0}_{b_1}\ket{0}_{b_2} \ , \hspace{0.1cm} \ket{1}_L=\ket{1}_{b_1}\ket{1}_{b_2} \ .
\end{gather}
Assume that we have a one-dimensional graph state with a qubit $a$ between two qubits $b_1,b_2$. We will show that measuring the operator $X_a$ results in $b_1,b_2$ becoming one logical qubit as in eq. (\ref{eq:LogicalQubit}). We denote their other neighbors by $c_1,c_2$ respectively, so the one-dimensional graph state has in its middle the qubits $c_1,b_1,a,b_2,c_2$ in this order. Denote by $\ket{\phi}$ the wave function of the graph, by $\ket{\phi_a}$ the wave function of the graph without the vertex $a$, and by $\ket{\phi_{b_1}},\ket{\phi_{b_2}}$ the wave functions of the following graphs; the graph obtained by erasing all the qubits to the right of $c_1$, and the graph obtained by erasing all the qubits to the left of $c_2$, respectively. By eq. (\ref{eq:RecursiveDefinitionOfGraphState}), one can write $\ket{\phi}$ in the following ways (up to normalizations):

\begin{gather}
    \label{eq:WritingTheWaveFunctionByDelitingab1b2GraphState}
    \ket{\phi}=\ket{0}_{a}\ket{\phi_{a}}+\ket{1}_a Z_{b_1}Z_{b_2}\ket{\phi_{a}} \ , \nonumber \\
    \ket{\phi_a}=\left(\ket{0}_{b_1}\ket{\phi_{b_1}}+\ket{1}_{b_1} Z_{c_1}\ket{\phi_{b_1}}\right)\left(\ket{0}_{b_2}\ket{\phi_{b_2}}+\ket{1}_{b_2} Z_{c_2}\ket{\phi_{b_2}}\right) \ .
\end{gather}
We now perform a $X_a$ measurement, which results in a projection onto one of the states (up to normalization) $\ket{0}_a\pm \ket{1}_a$. The resulting state reads (up to normalization):

\begin{gather}
    \ket{\psi}=\left(\bra{0}_a\pm \bra{1}_a\right)\ket{\phi}=\ket{0}_{b_1}\ket{0}_{b_2}\ket{\phi_{b_1}}\ket{\phi_{b_2}}\pm \ket{1}_{b_1}\ket{1}_{b_2}Z_{c_1}Z_{c_2}\ket{\phi_{b_1}}\ket{\phi_{b_2}} \ .
    \label{eq:GraphStateAfterApplyngXmeasurementOnQubita}
\end{gather}

If the sign between the terms in eq. (\ref{eq:GraphStateAfterApplyngXmeasurementOnQubita}) is minus then applying $X_{b_1}$ or $X_{b_2}$ on it will set the sign to plus, and by eq. (\ref{eq:RecursiveDefinitionOfGraphState}) and eq. (\ref{eq:LogicalQubit}) this is the desired creation of a logical qubit.

\subsection{Type-II Fusion}
\label{Subsec:Fusion Type-II}
Assume that we have two one-dimensional graph states we wish to fuse. Assume that one of those graph states already contains a logical qubit $L$ consisting of the qubits $a$ and $e$ (as in eq. (\ref{eq:LogicalQubit})), and choose a qubit $b$ from the other graph. By eq. (\ref{eq:RecursiveDefinitionOfGraphState}), the total wave function is:

\begin{gather}
    \label{eq:StartingWaveFunctionFusionTypeIIGraphStates}
    \ket{\psi}=\left(\ket{0}_{a}\ket{0}_{e}\ket{\phi_L}+\ket{1}_a\ket{1}_{e}\prod_{g\in n(L)}Z_{g}\ket{\phi_L}\right)\left(\ket{0}_{b}\ket{\phi_b}+\ket{1}_b\prod_{f\in n(b)}Z_{f}\ket{\phi_b}\right) \ .
\end{gather}

The Type-II fusion process is also a probabilistic gate as Type-I fusion, that involves erasing the qubits $a,b$, but without generating a new qubit --- instead, when succeeding, the qubit $e$ inherits the neighbors of $b$ (together with the neighbors of $L$ that became $e$). 
In the case of success (which has probability one-half that can be increased by using ancilla qubits, as for Type-I fusion), one of the following operators is applied to the state (\ref{eq:StartingWaveFunctionFusionTypeIIGraphStates}) (up to normalization):

{
\begin{gather}
    \label{eq:OperatorFusionTypeIISuccess}
    \bra{0}_a\bra{0}_b+ \bra{1}_a\bra{1}_b \ , \hspace{0.1cm} \bra{0}_a\bra{1}_b+ \bra{1}_a\bra{0}_b \ ,
\end{gather}
yielding one of the states:
\begin{gather}
    \ket{\psi_{new}}=\frac{1}{\sqrt{2}}\left(\ket{0}_{e}\ket{\phi_L}\ket{\phi_b}+ \ket{1}_e\prod_{g\in n(L)}Z_{g}\prod_{f\in n(b)}Z_{f}\ket{\phi_L}\ket{\phi_b}\right) \ ,
    \nonumber \\
    \ket{\psi_{new}}=\frac{1}{\sqrt{2}}\left(\ket{0}_{e}\prod_{f\in n(b)}Z_{f}\ket{\phi_L}\ket{\phi_b}+ \ket{1}_e\prod_{g\in n(L)}Z_{g}\ket{\phi_L}\ket{\phi_b}\right) \ .
\end{gather}
If we got the second option, one can apply the operator multiplication $\prod_{f\in n(b)}Z_f$ 
to get the first option, which by eq. (\ref{eq:RecursiveDefinitionOfGraphState}) is the desired fused graph state.
}

\if{\begin{gather}
    \label{eq:OperatorFusionTypeIISuccess}
    \bra{0}_a\bra{0}_b\pm \bra{1}_a\bra{1}_b \ ,
\end{gather}
yielding the state:

\begin{gather}
    \ket{\psi_{new}}=\frac{1}{\sqrt{2}}\left(\ket{0}_{e}\ket{\phi_L}\ket{\phi_b}\pm \ket{1}_e\prod_{g\in n(L)}Z_{g}\prod_{f\in n(b)}Z_{f}\ket{\phi_L}\ket{\phi_b}\right) \ .
\end{gather}
If the sign is negative, one can apply a $Z_e$ operator to change the sign to plus, so in both cases we end up with the plus sign, which by eq. (\ref{eq:RecursiveDefinitionOfGraphState}) is the desired fused graph state.}\fi

In the case of failure (which has probability one-half that can be lowered by using ancilla qubits, as for Type-I fusion), one of the following operators is applied on the state (\ref{eq:StartingWaveFunctionFusionTypeIIGraphStates}) (up to normalization):

\begin{gather}
    \label{eq:OperatorFusionTypeIIFailure}
    \left(\bra{0}_a\pm \bra{1}_a\right)\left(\bra{0}_b\mp \bra{1}_b\right) \ ,
\end{gather}
which results in erasing $a,b$, while keeping the original graph states separated. One may think of it as measuring $X_a,X_b$, with the constraint that the resulting eigenvalues must differ. On the first graph state, "measuring $X_a$" replaces the logical qubit by $e$, up to applying $Z_e$ in the case of a minus sign, so this leaves us with a regular one-dimensional graph state. On the second graph state, "measuring $X_b$" creates a logical qubit in this graph state, thus, a new Type-II fusion attempt can be made without the need to create a logical qubit.

Type-II fusion can also be described in the language of $f_1,f_2,f_3,f_4$ introduced in subsection \ref{Subsec:Fusion Type-I}, where now $f_1=\ket{0}_{e}\ket{\phi_L}$, $f_2=\ket{1}_{e}\prod_{g\in n(L)}Z_{g}\ket{\phi_L}$ and $f_3=\ket{\phi_b}$, $f_4=\prod_{f\in n(b)}Z_{f}\ket{\phi_b}$. The wave function (\ref{eq:StartingWaveFunctionFusionTypeIIGraphStates}) is again described by the form (\ref{eq:StartingWaveFunctionFusionTypeIGraphStatesInf1f2f3f4Language}). We now perform a unitary transformation on $Span\{{a_H^{\dagger},a_V^{\dagger},b_H^{\dagger},b_V^{\dagger}\}}$, yielding a new basis $c_H^{\dagger},c_V^{\dagger},d_H^{\dagger},d_V^{\dagger}$, and measure the channels $c_H^{\dagger},c_V^{\dagger},d_H^{\dagger},d_V^{\dagger}$.
The probabilities can be computed in the same manner as in Subsection \ref{Subsec:Fusion Type-I}, but with a different resulting wave function (than the wave function in eq. (\ref{eq:StartingWaveFunctionFusionTypeIGraphStatesInf1f2f3f4LanguageAfterTransferingToCandD})) before the measurements, resulting from transferring the channels $a,b$ through a diagonal polarizing beam splitter (see \cite{Kok_2009} for more information). 
{\color{black}Note that, as long as $f_1,f_2,f_3,f_4$ lie outside $Span\{{a_H^{\dagger},a_V^{\dagger},b_H^{\dagger},b_V^{\dagger}\}}$, $f_1,f_2$ orthogonal and of equal norm, and $f_3,f_4$ of equal norm, the computation of the probabilities in the event of success does not depend on whether $f_3,f_4$ are orthogonal. This is because in the case of success, the projected expression is proportional to $f_1f_3+f_2f_4$ or $f_1f_4+f_2f_3$, whose norms remain the same even if $f_3,f_4$ are not orthogonal. In the case of failure, the projected expressions are proportional to $(f_1+f_2)(f_3-f_4)$ or $(f_1-f_2)(f_3+f_4)$, and may change their norms, while the total probability of failure remains the same.}

\if{The important thing to note is that, as in the Type-I fusion case, the computation of the probabilities is not dependent on the specific forms of $f_1,f_2,f_3,f_4$, but only on them not belonging to $Span\{{a_H^{\dagger},a_V^{\dagger},b_H^{\dagger},b_V^{\dagger}\}}$.}\fi

\subsection{Generalized Type-II Fusion}
\label{subsec:Generalized Type-II Fusion}
One can consider a general unitary matrix that transfers between the modes $a_H^{\dagger},a_V^{\dagger},b_H^{\dagger},b_V^{\dagger}$ and $c_H^{\dagger},c_V^{\dagger},d_H^{\dagger},d_V^{\dagger}$:

\begin{gather}
        \begin{bmatrix}
        a^\dag_H \\
        a^\dag_V \\
        b^\dag_H \\
        b^\dag_V
    \end{bmatrix} = \begin{bmatrix}
     U_{11} & U_{12} & U_{13} & U_{14} \\
     U_{21} & U_{22} & U_{23} & U_{24} \\
     U_{31} & U_{32} & U_{33} & U_{34} \\
     U_{41} & U_{42} & U_{43} & U_{44}
    \end{bmatrix}
    \begin{bmatrix}
        c^\dag_H \\
        c^\dag_V \\
        d^\dag_H \\
        d^\dag_V
    \end{bmatrix} \ .
    \label{eq:Unitary transformation}
\end{gather}

After measuring the channels $c_H^{\dagger},c_V^{\dagger},d_H^{\dagger},d_V^{\dagger}$, which we denote by $i=1,2,3,4$, we detect two photons, either both in the same channel or one in each of two different channels. If the two photons are detected in the same channel, $(i,i)$, then the resulting state is \cite{rimock2024generalizedtypeiifusion}:
\begin{gather}
    \ket{\phi}_{ii}=\frac{1}{N_{ii}}(U_{1i}f_{1}+U_{2i}f_{2})(U_{3i}f_{3}+U_{4i}f_{4}) \ ,
    \label{eq:Wave function of non-relevant state}
\end{gather}
which is a product state that is called a "non-relevant state". The probability for this state is:

\begin{gather}
        p_{ii}=\left(|U_{1i}|^2+|U_{2i}|^2\right)\left(|U_{3i}|^2+|U_{4i}|^2\right) \ .
        \label{eq:Probability of non-relevant state}
\end{gather}

If the two photons are detected in two different channels, $(i,j)$, then the resulting state is:

\begin{gather}
    \ket{\phi}_{ij}=\frac{1}{N_{ij}}(a_{ij}f_{1}f_{3}+b_{ij}f_{1}f_{4}+c_{ij}f_{2}f_{3}+d_{ij}f_{2}f_{4})=
    \begin{bmatrix}
        f_1 & f_2
    \end{bmatrix}
    M_{ij}
    \begin{bmatrix}
        f_3 \\
        f_4
    \end{bmatrix} \nonumber \\
    M_{ij}=\frac{1}{N_{ij}}\begin{bmatrix}
        a_{ij} & b_{ij} \\
        c_{ij} & d_{ij}
    \end{bmatrix} \ ,
    \label{eq:Wave function of relevant state}
\end{gather}
where the coefficients are: 
\begin{gather}
    a_{ij}=U_{1i}U_{3j}+U_{1j}U_{3i}
    \hspace{0.5cm}
    b_{ij}=U_{1i}U_{4j}+U_{1j}U_{4i} \nonumber\\
    c_{ij}=U_{2i}U_{3j}+U_{2j}U_{3i}
    \hspace{0.5cm}
    d_{ij}=U_{2i}U_{4j}+U_{2j}U_{4i} \nonumber\\
    N_{ij}=\sqrt{|a_{ij}|^2+|b_{ij}|^2+|c_{ij}|^2+|d_{ij}|^2}=\sqrt{4p_{ij}} \ ,
    \label{eq:The coefficients of the wave function}
\end{gather}
and $p_{ij}$ is the probability of getting the state.

For each of these final states, the "quality" measure introduced in \cite{rimock2024generalizedtypeiifusion} is the entanglement entropy that arises when dividing the Hilbert space containing the total wave function into the Hilbert spaces of the surviving qubits of the two original graphs. Because $f_1,f_2$ and $f_3,f_4$ are orthonormal bases, this entanglement entropy is also the entanglement entropy of the 2-qubit $a,b$ state onto which we project. This entanglement entropy can be computed by evaluating the reduced density matrix after tracing out $Span\{f_3,f_4\}$ for the state (\ref{eq:Wave function of relevant state}), which is:

\begin{gather}
    \rho_{ij}=M_{ij}M_{ij}^{\dag}=\frac{1}{N_{ij}^{2}}
    \left(
    \begin{bmatrix}
            |a_{ij}|^2+|b_{ij}|^2 & a_{ij}c_{ij}^*+b_{ij}d_{ij}^* \\ a_{ij}^*c_{ij}+b_{ij}^*d_{ij} & |c_{ij}|^2+|d_{ij}|^2
    \end{bmatrix}
    \right) \ .
    \label{eq:Reduced Density matrix Graph state}
\end{gather}
From eq. (\ref{eq:Reduced Density matrix Graph state}) one can compute the determinant:

\begin{gather}
    det(\rho_{ij})=\left(det\left(M_{ij}\right)\right)^2=\frac{|a_{ij}d_{ij}-b_{ij}c_{ij}|^2}{N_{ij}^4}\nonumber\\=\left|\frac{(U_{1i}U_{2j}-U_{1j}U_{2i})(U_{3j}U_{4i}-U_{3i}U_{4j})}{4p_{ij}}\right|^2 \ ,
    \label{eq:Determinant identities Graph states}
\end{gather}
from which the eigenvalues of the reduced density matrix and the entanglement entropy read:

\begin{gather}
    \lambda_{ij}=\frac{1+\sqrt{1-4\det(\rho_{ij})}}{2} \nonumber \\
    1-\lambda_{ij}=\frac{1-\sqrt{1-4\det(\rho_{ij})}}{2} \nonumber \\
    S_{ij}=-\lambda_{ij} \log_2 \lambda_{ij} -(1-\lambda_{ij} ) \log_2 {(1-\lambda_{ij})} \ .
    \label{eq:Entropy and eigenvalues of the reduced density matrix Graph states}
\end{gather}

In fact, all the equations (\ref{eq:Wave function of non-relevant state}),(\ref{eq:Probability of non-relevant state}),(\ref{eq:Wave function of relevant state}),(\ref{eq:The coefficients of the wave function}),(\ref{eq:Reduced Density matrix Graph state}),(\ref{eq:Determinant identities Graph states}) and (\ref{eq:Entropy and eigenvalues of the reduced density matrix Graph states}) still hold where the transformation in eq. (\ref{eq:Unitary transformation}) is more general and includes vacuum ancilla qubits, meaning that we transform the four channels $a^\dag_H,a^\dag_V,b^\dag_H,b^\dag_V$ to $N$ channels by an isometry, which can be described by an $N$ by $N$ unitary matrix:

\begin{gather}
        \begin{bmatrix}
        a^\dag_H \\
        a^\dag_V \\
        b^\dag_H \\
        b^\dag_V \\
        \ket{vac}_1 \\
        . \\
        . \\
        . \\
        \ket{vac}_N
    \end{bmatrix} = U
    \begin{bmatrix}
        c^\dag_1 \\
        c^\dag_2 \\
        c^\dag_3 \\
        c^\dag_4 \\
        . \\
        . \\
        . \\
        c^\dag_N
    \end{bmatrix} \ ,
    \label{eq:Unitary transformation with ancilla}
\end{gather}
and the only difference is that, when referring to $(i,j)$ or $(i,i)$ states, $i$ and $j$ are between $1$ and $N$ instead of $1$ and $4$.

{\color{black}The resulting state (\ref{eq:Wave function of relevant state}) is maximally entangled iff $\rho_{ij}=\frac{1}{2}Id$, meaning iff $M_{ij}$ is unitary up to multiplication by $\frac{1}{\sqrt{2}}$.}

\subsection{Projecting onto a non-Bell state}
\label{Subsec:Projecting onto non-Bell state}
The resulting state (\ref{eq:Wave function of relevant state}) from the described fusion process is a special case of acting on the initial wave function (\ref{eq:StartingWaveFunctionFusionTypeIIGraphStates}) with a general projection onto a two-qubit state of $a,b$:

\begin{gather}
    \label{eq:projectionOperatorForGeneralU}
    A\bra{0}_a\bra{0}_b+B\bra{0}_a\bra{1}_b+C\bra{1}_a\bra{0}_b+D\bra{1}_a\bra{1}_b \ ,
\end{gather}
which results in the final state:

\begin{gather}
    A\ket{0}_e\ket{\phi_L}
    \ket{\phi_b}
    +B\ket{0}_e\ket{\phi_L}\prod_{f\in n(b)} Z_f\ket{\phi_b}
    +C\ket{1}_e\prod_{g\in n(L)} Z_g \ket{\phi_L} \ket{\phi_b}
    \nonumber\\
    +D\ket{1}_e\prod_{g\in n(L)} Z_g \ket{\phi_L}\prod_{f\in n(b)} Z_f \ket{\phi_b} \ .
    \label{eq:Cluster state after measuring a,b for general unitary transformation}
\end{gather}

In \cite{rimock2024generalizedtypeiifusion}, some of the final states (\ref{eq:Cluster state after measuring a,b for general unitary transformation}) were classified as final states that are the desired fused graph state up to single-qubit rotations, and weighted graph states up to single-qubit rotations.
If $|n(b)|=2$, which means that $b$ is not at the endpoint of its original graph, then if $|A|=|D|$ and $B=C=0$, or $|B|=|C|$ and $A=D=0$, then the resulting state is the required fused graph state up to $Z$ rotation on the qubit $e$.
If $|n(b)|=1$, which means that $b$ is at the endpoint of its original graph, then if the previous conditions hold, the resulting state is again the required fused graph state up to $Z$ rotation on the qubit $e$. But, if we also allow single-qubit rotations on the single qubit in $n(b)$ (on the resulting state), then there are more $A,B,C,D$ that provide the required fused graph state up to single-qubit rotations, and they can be parameterized as follows:

\begin{gather}
        A,D=\frac{e^{i\theta_{1,2}}}{\sqrt{2}}\cos{\phi}  \hspace{0.5cm}B,C=\frac{ie^{i\theta_{1,2}}}{\sqrt{2}}\sin{\phi} 
        \ .
        \label{eq:General form of A,B,C,D in order to get cluster state after hard transformations}
\end{gather}
These are also sub-cases of all the $A,B,C,D$ yielding a final state that is a weighted graph state up to single-qubit rotations, which can be parameterized as follows:

\begin{gather}
        (A,B)=\frac{e^{i\theta_1}}{\sqrt{2}}(\cos{\phi_1}, i\sin{\phi_1}), \nonumber \\
        (C,D)=\frac{e^{i\theta_2}}{\sqrt{2}}(i\sin{\phi_2},\cos{\phi_2}) \ ,
        \label{eq:Parameterization of A,B,C,D in the case of graph state}
\end{gather}
with the resulting weight:

 \begin{gather}
        \chi=2(\phi_1-\phi_2)+\pi \ .
        \label{eq:chi by phi1,2}
\end{gather} 
Equations (\ref{eq:Parameterization of A,B,C,D in the case of graph state}) provide a parametrization of the solutions to the following equations:

\begin{gather}
        |A|^2+|B|^2=|C|^2+|D|^2=\frac{1}{2} \ , \nonumber \\
        \Re{AB^*}=\Re{CD^*}=0 \ .
        \label{Conditions on A,B,C,D for graph state}
\end{gather}

\section{Fusions for Weighted Graph States}
\label{Section:Fusions for weighted graph state}
In this section we detail our findings on the use of Type-I and Type-II fusions for weighted graph states. While the Type-I fusion works the same for weighted graph states, the Type-II fusion faces two obstacles: i) creating the logical qubit is not possible for every one-dimensional weighted graph state, but only for certain weighted graph states. ii) While in the case of success, Type-II fusion works well for weighted graph states (given that the logical qubit was created), in the case of failure, Type-II is problematic, requiring the use of boosting to make the probability of failure negligible. Those problems can be avoided by applying the regular Type-II fusion to 1-dim weighted graph states that are padded with $\pi$-weights, at the cost of many additional qubits and Type-I fusion operations. In addition, we show ways to create a pool of 2-qubit weighted graph states required as building blocks for generating larger weighted graph states by fusion processes. We also analyze the result of a non-Bell projection and generalized fusion on weighted graph states, compute {\color{black}the probability and} the resulting entanglement entropy, and classify the resulting state. 

\subsection{Type-I fusion for weighted graph states}

\label{Subsec:Fusion type-I for weighted graph state}
\begin{figure}
    \centering
    \includegraphics[width=0.9\linewidth]{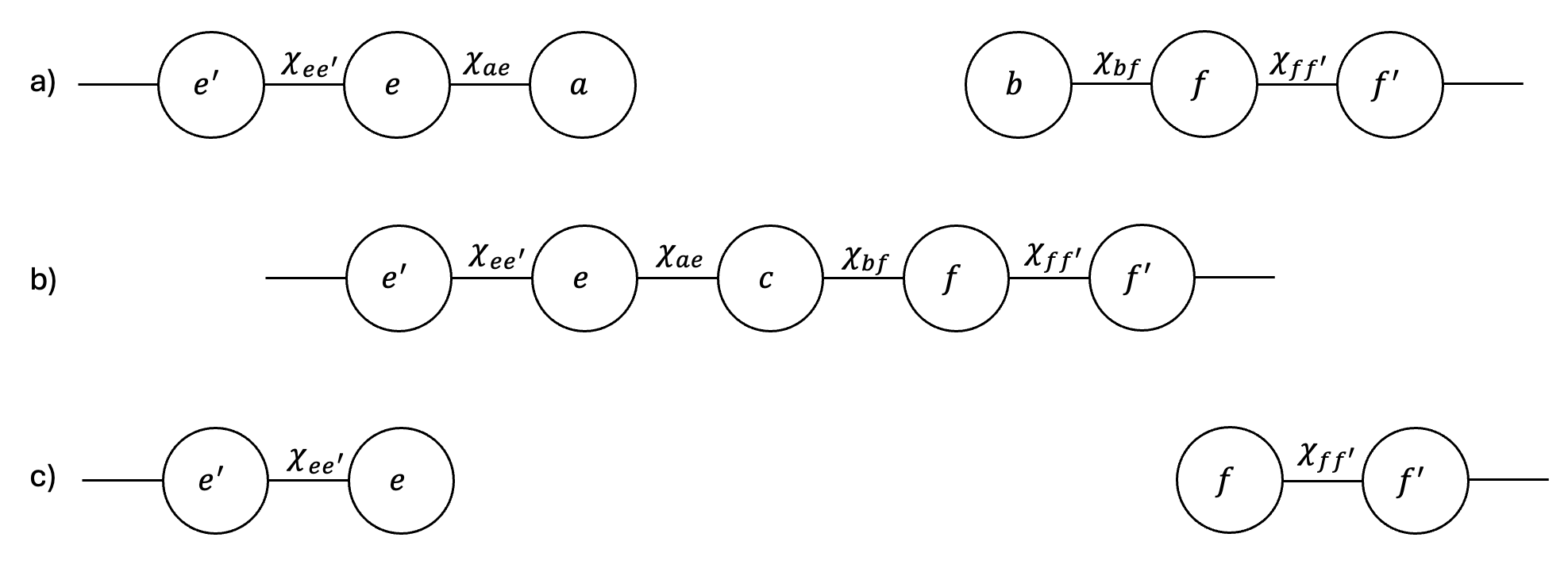}
    \caption{Type-I fusion of two 1-dim weighted graph states. We begin with two separated 1-dim weighted graph states (a), and apply Type-I fusion gate on the qubits $a$ and $b$. In the case of successful fusion, we get the desired fused weighted graph state (b). In the case of failure of the fusion, the original weighted graph states remain separated, and each loses one qubit (c).}
    \label{fig:FusionTypeIOfWeightedGraphStates}
\end{figure}

In this subsection we show that Type-I fusion also works for one-dimensional weighted graph states, by repeating all the steps of subsection \ref{Subsec:Fusion Type-I}. We begin with two one-dimensional weighted graph states with qubits $a,b$ located at the endpoints of these weighted graphs. We denote by $\chi_{ae},\chi_{bf}$ the weights of the edges of $a$ and $b$, where $e,f$ are the neighbors of $a$ and $b$, respectively (see figure \ref{fig:FusionTypeIOfWeightedGraphStates}.a). By eq. (\ref{eq:RecursiveDefinitionOfWeightedGraphState}), the complete wave function can be written as:

\begin{gather}
    \ket{\psi}=\left(\ket{0}_{a}\ket{\phi_a}+\ket{1}_a\prod_{e\in n(a)}e^{-i\chi_{ae}\ket{1}_e\bra{1}_e}\ket{\phi_a}\right)\cdot\nonumber \\
    \cdot\left(\ket{0}_{b}\ket{\phi_b}+\ket{1}_b\prod_{f\in n(b)}e^{-i\chi_{bf}\ket{1}_f\bra{1}_f}\ket{\phi_b}\right) \ .
    \label{eq:StartingWaveFunctionFusionTypeIWeightedGraphStates}
\end{gather}

We now apply Type-I fusion. In the case of success, we operate with (\ref{eq:OperatorFusionTypeISuccess}) on the state (\ref{eq:StartingWaveFunctionFusionTypeIWeightedGraphStates}) and get the state:

\begin{gather}
    \ket{\psi_{new}}=\frac{1}{\sqrt{2}}\left(\ket{0}_{c}\ket{\phi_a}\ket{\phi_b}\pm \ket{1}_c\prod_{e\in n(a)}e^{-i\chi_{ae}\ket{1}_e\bra{1}_e}\prod_{f\in n(b)}e^{-i\chi_{bf}\ket{1}_f\bra{1}_f}\ket{\phi_a}\ket{\phi_b}\right) \ .
    \label{eq:TypeIFusionOFWeightedGraphStatesSuccessWaveFunction}
\end{gather}
If the sign is minus one can apply $Z_c$ operator to change the sign to plus, so in both cases we end up with the sign plus which by eq. (\ref{eq:RecursiveDefinitionOfWeightedGraphState}) is the desired fused weighted graph state, where we replace $a,b$ with $c$ that inherits the neighbors and weights of $a,b$ {(see figure \ref{fig:FusionTypeIOfWeightedGraphStates}.b)}.
In the case of failure, we apply one of the operators (\ref{eq:OperatorFusionTypeIIFailure}), which results in $Z$ measurement of the qubits $a,b$. As explained in eq. (\ref{eq:WeightedGraphStateAfterZmeasurement}), this will result in erasing qubits $a,b$ and their edges {(see figure \ref{fig:FusionTypeIOfWeightedGraphStates}.c)}, just as in the graph state case, hence the failure case is also identical for graph states and weighted graph states.

The computations of the probabilities of success and of failure are still the same (one-half without using ancilla qubits), but this requires an explanation. There is a major difference between the cases of graph states and weighted graph states; for graph states, when we write the recursive definition (\ref{eq:RecursiveDefinitionOfGraphState}), $\ket{0}_a,\ket{1}_a$ are multiplied by $
\ket{\phi_a}$,$\prod_{e\in n(a)}Z_{e}\ket{\phi_a}$, that are orthogonal to each other and have the same norm. 
{For weighted graph states, 
$
\ket{\phi_a}$,$\prod_{e\in n(a)}e^{-i\chi_{ae}\ket{1}_e\bra{1}_e}\ket{\phi_a}$ are not orthogonal
(but have the same norm).} Returning to the language of $f_1,f_2,f_3,f_4$ represented in subsections \ref{Subsec:Fusion Type-I} and \ref{Subsec:Fusion Type-II}, we can set $f_1=\ket{\phi_a}$,$f_2=\prod_{e\in n(a)}e^{-i\chi_{ae}\ket{1}_e\bra{1}_e}\ket{\phi_a}$ and $f_3=\ket{\phi_b}$,$f_4=\prod_{f\in n(b)}e^{-i\chi_{bf}\ket{1}_f\bra{1}_f}\ket{\phi_b}$, and the initial wave function (\ref{eq:StartingWaveFunctionFusionTypeIWeightedGraphStates}) becomes of the form (\ref{eq:StartingWaveFunctionFusionTypeIGraphStatesInf1f2f3f4Language}). The fusion process is as in eq. (\ref{eq:StartingWaveFunctionFusionTypeIGraphStatesInf1f2f3f4LanguageAfterTransferingToCandD}), but now we have the couples $f_1,f_2$ and $f_3,f_4$ being not orthogonal {\color{black}(but have the same norm)}. However, as discussed in subsection \ref{Subsec:Fusion Type-I}, the computation of the probabilities still holds even when $f_1,f_2$ and $f_3,f_4$ are not orthonormal couples, hence the probabilities remain the same.

In the graph state case, the creation of one-dimensional graph states required a "pool" of Bell pairs. In our case, the analogous required pairs are 2-qubit weighted graph states. In subsection \ref{Subsec:Creating 2-qubit weighted graph states from GHZ states} we show that one can create these pairs by measurements on a 3-qubit one-dimensional graph state (with success probability of 100 percent for any target weight), or even on 3-qubit one-dimensional weighted graph state (with success probability of 100 percent for a range of target weights). One can also create these weighted pairs from Bell pairs. For instance, in \cite{rimock2024generalizedtypeiifusion} it was shown that by applying generalized Type-II fusion on two one-dimensional graph states, one can create a weighted graph state with a target weight $\chi$ on one of the edges, while all the other edges have weight $\pi$ as regular graph state. Then, one can apply $Z$ measurements to all the qubits except the two with the edge with weight $\chi$, and as will be shown by eq. (\ref{eq:WeightedGraphStateAfterZmeasurement}), this will result in the desired two-qubit weighted graph state with weight $\chi$. In fact, when the two one-dimensional graph states are a single logical qubit and a two-qubit graph state, after the appropriate generalized fusion the resulting state is already a 2-qubit weighted graph state, without the need to apply $Z$ measurement.

\begin{figure}
    \centering
    \includegraphics[width=0.7\linewidth]{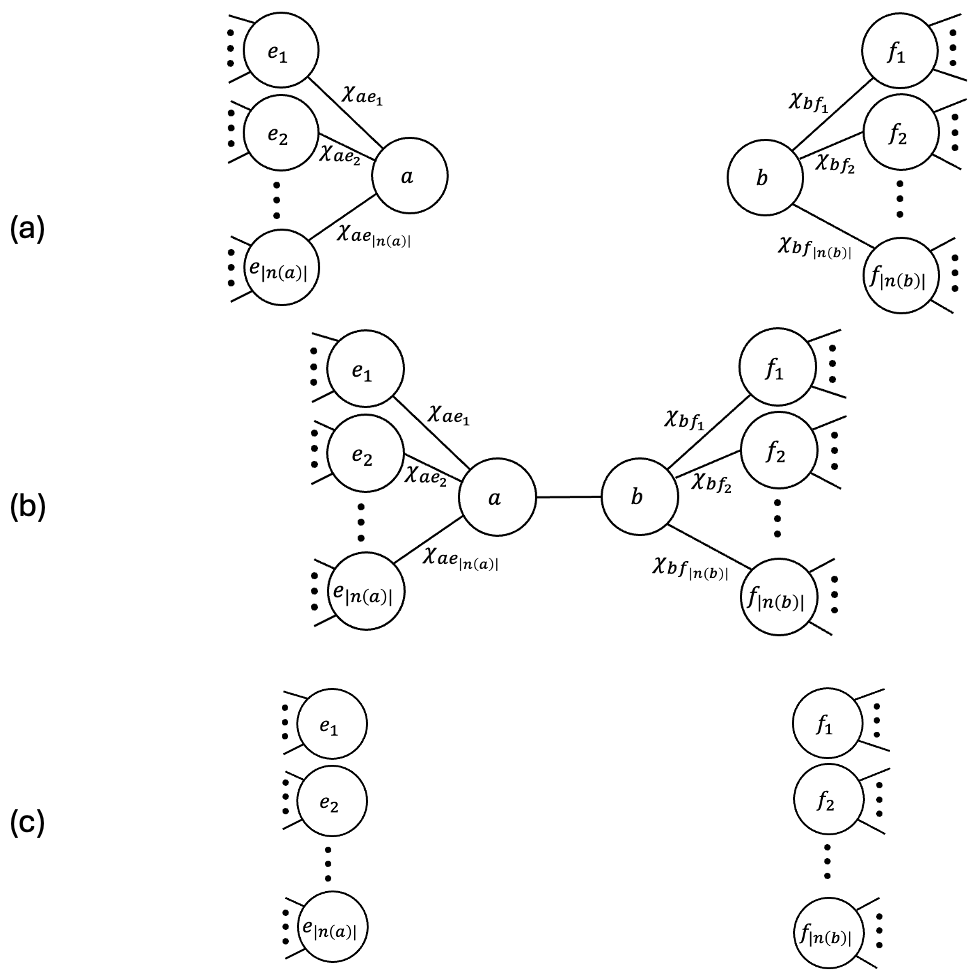}
    \caption{Type-I fusion of two general weighted graph states. We begin with two separated weighted graph states (a), and apply Type-I fusion gate on the qubits $a$ and $b$. In the case of successful fusion, we get the desired fused weighted graph state (b). In the case of failure of the fusion, the original weighted graph states not only remain separated and each loses one qubit, but also lose some of their structure, exactly like in the graph states case (c).}
    \label{fig:FusionTypeIOfWeightedGraphStatesGeneral}
\end{figure}

{Finally, it is worth noting that by equations (\ref{eq:TypeIFusionOFWeightedGraphStatesSuccessWaveFunction}) and (\ref{eq:WeightedGraphStateAfterZmeasurement}), Type-I fusion will work the same when the original weighted graph states are not 1-dim (see figure \ref{fig:FusionTypeIOfWeightedGraphStatesGeneral}.a), and a successful fusion will replace the qubits $a,b$ with a qubit $c$ that has their original neighbors (see figure \ref{fig:FusionTypeIOfWeightedGraphStatesGeneral}.b). But, in the case of failure, the structure of the original weighted graph states will be damaged (see figure \ref{fig:FusionTypeIOfWeightedGraphStatesGeneral}.c), making this fusion protocol unattractive for fusing weighted graph states from not leaves qubits, just like in the graph states case.  
}

\subsection{Creating a logical qubit for one-dimensional weighted graph states}
\label{Subsec:Creating a logical qubit for 1-dim weighted graph state}
It is a well-known fact that when applying a $Z$ measurement on a vertex in a graph state, this operation erases this vertex and its edges from the graph. This is true also for weighted graph states \cite{PhysRevA.110.022431} --- one can apply a $Z$ measurement on the state (\ref{eq:RecursiveDefinitionOfWeightedGraphState}), which will leave us with one of the states:

\begin{gather}
    \label{eq:WeightedGraphStateAfterZmeasurement}
    \ket{\phi} \hspace{0.1cm} \text{or} \prod_{b\in n(a)}e^{-i\chi_{ab}\ket{1}_b\bra{1}_b}\ket{\phi} \ ,
\end{gather}
both are states associated with the weighted graph obtained by erasing the vertex $a$ and its edges, up to single-qubit rotations.

The same does not hold for $X$ or $Y$ measurements on weighted graph states, and $X$ measurement is an essential tool in order to create a logical qubit {(as in eq. (\ref{eq:LogicalQubit}))}, that is used in the Type-II fusion process. In a recent paper \cite{yamazaki2025measurementbasedquantumcomputationweighted}, the concept of weighted Pauli measurement was introduced, and it was shown that given a one-dimensional graph with equal weights or
weights of opposite signs, X-like measurements are possible. 
We arrive in this subsection at the same conclusion, and we also show that these are the only possible ways to do so.

\begin{figure}
    \centering
    \includegraphics[width=0.5\linewidth]{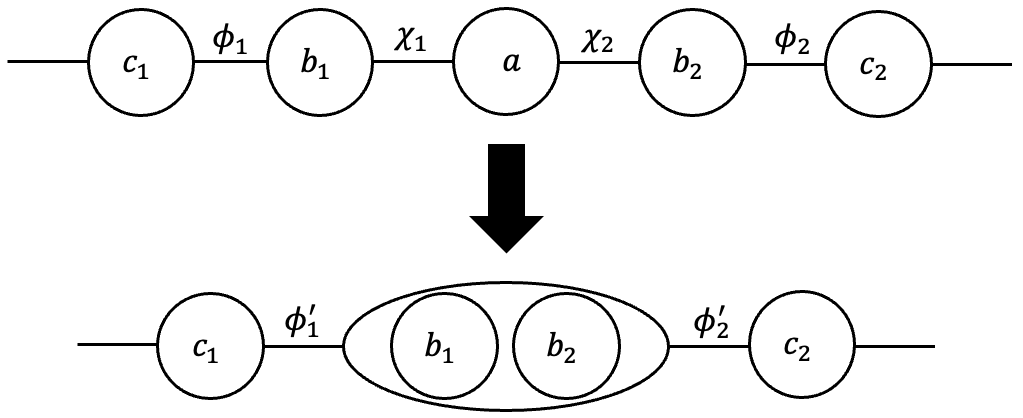}
    \caption{{X-like measurement on weighted graph states; the goal is to measure the middle qubit $a$, in such a way that its neighbors $b_1,b_2$ will become a logical qubit. This is possible in the two cases, $\chi_1=\chi_2=\chi$ and $\chi_1=2\pi-\chi_2=\chi$, and in both cases the appropriate measurement works only in one of its results --- with success probability as in equation (\ref{eq:probability of X-like measurement to succeed}). In the case $\chi_1=2\pi-\chi_2=\chi$, the weights $\phi_1,\phi_2$ between $b_1,b_2$ and their neighbors $c_1,c_2$ that are not $a$, will change --- the sign of one of them will be flipped.}}
    \label{fig:XmeasurementWeightedGraphState}
\end{figure}

Assume that we have a one-dimensional weighted graph state, and we have a qubit $a$ we wish to measure in order for its adjacent qubits $b_1,b_2$ to become one logical qubit. We denote their other neighbors by $c_1,c_2$ respectively, so the one-dimensional weighted graph state has in its middle the qubits $c_1,b_1,a,b_2,c_2$ in this order. Denote by $\chi_1,\chi_2$ the weights between $a,b_1$ and $a,b_2$, respectively, and by $\phi_1,\phi_2$ the weights between $b_1,c_1$ and $b_2,c_2$, respectively (see figure \ref{fig:XmeasurementWeightedGraphState}). Denote by $\ket{\phi}$ the wave function of the weighted graph, by $\ket{\phi_a}$ the wave function of the weighted graph without the vertex $a$, and by $\ket{\phi_{b_1}},\ket{\phi_{b_2}}$ the wave functions of the following weighted graphs; the weighted graph obtained by erasing all the qubits to the right of $c_1$, and the weighted graph obtained by erasing all the qubits to the left of $c_2$, respectively. By eq. (\ref{eq:RecursiveDefinitionOfWeightedGraphState}), one can write $\ket{\phi}$ in the following ways (up to normalizations):

\begin{gather}
    \ket{\phi}=\ket{0}_{a}\ket{\phi_{a}}+\ket{1}_a e^{-i\chi_{1}\ket{1}_{b_1}\bra{1}_{b_1}}e^{-i\chi_{2}\ket{1}_{b_2}\bra{1}_{b_2}}\ket{\phi_{a}} \ , \nonumber \\
    \ket{\phi_a}=\left(\ket{0}_{b_1}\ket{\phi_{b_1}}+\ket{1}_{b_1} e^{-i\phi_{1}\ket{1}_{c_1}\bra{1}_{c_1}}\ket{\phi_{b_1}}\right)\left(\ket{0}_{b_2}\ket{\phi_{b_2}}+\ket{1}_{b_2} e^{-i\phi_{2}\ket{1}_{c_2}\bra{1}_{c_2}}\ket{\phi_{b_2}}\right) \ .
    \label{eq:WritingTheWaveFunctionByDelitingab1b2}
\end{gather}

Assume that we now apply some measurement on qubit $a$ that projects it onto the state $A\ket{0}_a+B\ket{1}_a$, where $A,B$ are complex numbers with $|A|^2+|B|^2=1$ for normalization (note that $A,B\ne 0$ or else we would have made a $Z$ measurement). The resulting state reads (up to normalization):

\begin{gather}
    \ket{\psi}=\left(A\bra{0}_a+B\bra{1}_a\right)\ket{\phi}=\left(A+B\right)\ket{0}_{b_1}\ket{0}_{b_2}\ket{\phi_{b_1}}\ket{\phi_{b_2}}+ \nonumber \\
    +\left(A+Be^{-i\chi_1}\right)\ket{1}_{b_1}\ket{0}_{b_2}e^{-i\phi_{1}\ket{1}_{c_1}\bra{1}_{c_1}}\ket{\phi_{b_1}}\ket{\phi_{b_2}}+\nonumber \\
    +\left(A+Be^{-i\chi_2}\right)\ket{0}_{b_1}\ket{1}_{b_2}e^{-i\phi_{2}\ket{1}_{c_2}\bra{1}_{c_2}}\ket{\phi_{b_1}}\ket{\phi_{b_2}}+\nonumber \\
    +\left(A+Be^{-i\left(\chi_1+\chi_2\right)}\right)\ket{1}_{b_1}\ket{1}_{b_2}e^{-i\phi_{1}\ket{1}_{c_1}\bra{1}_{c_1}}e^{-i\phi_{2}\ket{1}_{c_2}\bra{1}_{c_2}}\ket{\phi_{b_1}}\ket{\phi_{b_2}} \ .
    \label{eq:WeightedGraphStateAfterApplyngGeneralProjectionOnQubita}
\end{gather}

In order to get our desired logical qubit (\ref{eq:LogicalQubit}), by eq. (\ref{eq:RecursiveDefinitionOfWeightedGraphState}) we want to get the state (up to normalization):

\begin{gather}
    \ket{0}_{b_1}\ket{0}_{b_2}\ket{\phi_{b_1}}\ket{\phi_{b_2}}+\ket{1}_{b_1}\ket{1}_{b_2}e^{-i\phi_{1,new}\ket{1}_{c_1}\bra{1}_{c_1}}e^{-i\phi_{2,new}\ket{1}_{c_2}\bra{1}_{c_2}}\ket{\phi_{b_1}}\ket{\phi_{b_2}} \ .
    \label{eq:DesiredWeightedGraphStateWithb1b2BeingLogicalQubit}
\end{gather}

{With respect to the bipartition of the total Hilbert space to the Hilbert space of $b_1,b_2$ and the Hilbert space of the other qubits, the desired state (\ref{eq:DesiredWeightedGraphStateWithb1b2BeingLogicalQubit}) has Schmidt rank $2$, while the state (\ref{eq:WeightedGraphStateAfterApplyngGeneralProjectionOnQubita}) has Schmidt rank as the number of non-zero coefficients between $A+B,A+Be^{-i\chi_1},A+Be^{-i\chi_2},A+Be^{-i\left(\chi_1+\chi_2\right)}$. That is because $\ket{0}_{b_1}\ket{0}_{b_2},\ket{1}_{b_1}\ket{0}_{b_2},\ket{0}_{b_1}\ket{1}_{b_2},\ket{1}_{b_1}\ket{1}_{b_2}$ are orthogonal to each other, and the states $\ket{\phi_{b_1}}\ket{\phi_{b_2}}$,$e^{-i\phi_{1}\ket{1}_{c_1}\bra{1}_{c_1}}\ket{\phi_{b_1}}\ket{\phi_{b_2}}$,$e^{-i\phi_{2}\ket{1}_{c_2}\bra{1}_{c_2}}\ket{\phi_{b_1}}\ket{\phi_{b_2}}$ and $e^{-i\phi_{1}\ket{1}_{c_1}\bra{1}_{c_1}}e^{-i\phi_{2}\ket{1}_{c_2}\bra{1}_{c_2}}\ket{\phi_{b_1}}\ket{\phi_{b_2}}$ are linear independent (see proof in appendix \ref{Appendix:Proof of linear independent}). The Schmidt rank is invariant under single-qubit rotations (and also under any rotations in the subspace of $b_1,b_2$ or the subspace of the other qubits), hence we must have two zeros between $A+B,A+Be^{-i\chi_1},A+Be^{-i\chi_2},A+Be^{-i\left(\chi_1+\chi_2\right)}$. Any choice of a couple of those terms that is not the first and fourth, or second and third, will result in $\chi_1=0$ or $\chi_2=0$, and we are left with two cases:}


\if{In appendix \ref{Appendix:ProofOfLogicalQubitWeightedGraphState} we prove that in order to get (\ref{eq:DesiredWeightedGraphStateWithb1b2BeingLogicalQubit}) by (\ref{eq:WeightedGraphStateAfterApplyngGeneralProjectionOnQubita}) using only single-qubit rotations on $b_1$,$b_2$, there are only two cases where there is a solution:}\fi

\textbf{Case 1} --- the second and third terms are zero. In this case one gets $A+Be^{-i\chi_1}=A+Be^{-i\chi_2}=0$, and this is possible only when the two weights $\chi_1,\chi_2$ are equal, $\chi_1=\chi_2=\chi$. Then one gets $B=-Ae^{i\chi}$ and the state (up to normalization):

\if{\textbf{case 1} --- The two weights $\chi_1,\chi_2$ are equal, $\chi_1=\chi_2=\chi$, and $B=-e^{i\chi}A$. The resulting state is (up to normalization):}\fi

\begin{gather}
    \left(1-e^{i\chi}\right)\ket{0}_{b_1}\ket{0}_{b_2}\ket{\phi_{b_1}}\ket{\phi_{b_2}}+\nonumber \\
    +\left(1-e^{-i\chi}\right)\ket{1}_{b_1}\ket{1}_{b_2}e^{-i\phi_{1}\ket{1}_{c_1}\bra{1}_{c_1}}e^{-i\phi_{2}\ket{1}_{c_2}\bra{1}_{c_2}}\ket{\phi_{b_1}}\ket{\phi_{b_2}} \ .
    \label{eq:WeightedGraphStateAfterApplyngGeneralProjectionOnQubitaWhereChi1EqualsChi2}
\end{gather}

Because $|1-e^{i\chi}|=|1-e^{-i\chi}|$, one can get the state (\ref{eq:DesiredWeightedGraphStateWithb1b2BeingLogicalQubit}) by a single-qubit rotation on either $b_1$ or $b_2$, specifically $e^{i\frac{\pi-\chi}{2} Z_{b_1}}$ or $e^{i\frac{\pi-\chi}{2} Z_{b_2}}$. 

\textbf{Case 2} --- the first and fourth terms are zero. In this case one gets $A+B=A+Be^{-i\left(\chi_1+\chi_2\right)}=0$, and this is possible only when the sum of the two weights $\chi_1,\chi_2$ is $\chi_1+\chi_2=2\pi$, $\chi_1=2\pi-\chi$ and $\chi_2=\chi$. Then one gets $B=-A$ and the state (up to normalization):

\if{\textbf{case 2} --- The sum of the two weights $\chi_1,\chi_2$ is $\chi_1+\chi_2=2\pi$, $\chi_1=2\pi-\chi$ and $\chi_2=\chi$, and $B=-A$. The resulting state is (up to normalization):}\fi

\begin{gather}
    \left(1-e^{i\chi}\right)\ket{1}_{b_1}\ket{0}_{b_2}e^{-i\phi_{1}\ket{1}_{c_1}\bra{1}_{c_1}}\ket{\phi_{b_1}}\ket{\phi_{b_2}}+\nonumber \\
    +\left(1-e^{-i\chi}\right)\ket{0}_{b_1}\ket{1}_{b_2}e^{-i\phi_{2}\ket{1}_{c_2}\bra{1}_{c_2}}\ket{\phi_{b_1}}\ket{\phi_{b_2}} \ .
    \label{eq:WeightedGraphStateAfterApplyngGeneralProjectionOnQubitaWhereChi1EqualsMinusChi2}
\end{gather}

By applying $X_{b_1}$ and $e^{i\phi_1 \ket{1}_{c_1}\bra{1}_{c_1}}$ on the state (\ref{eq:WeightedGraphStateAfterApplyngGeneralProjectionOnQubitaWhereChi1EqualsMinusChi2}) one gets the state (\ref{eq:WeightedGraphStateAfterApplyngGeneralProjectionOnQubitaWhereChi1EqualsChi2}) with the sign of $\phi_1$ flipped, and from there the desired wave function (\ref{eq:DesiredWeightedGraphStateWithb1b2BeingLogicalQubit}), again with the sign of $\phi_1$ flipped. One can also do the same for $b_2$ and $c_2$ instead and get the desired wave function (\ref{eq:DesiredWeightedGraphStateWithb1b2BeingLogicalQubit}) with the sign of $\phi_2$ flipped. 

{The option of getting the sign of $\phi_1$ flipped or the sign of $\phi_2$ flipped indicates that one can flip the sign of the weights on all the edges of a specific qubit, and also gives a way to do so. Given a vertex $v$ on a weighted graph, one can apply $X_v$ on $v$, and $e^{i\chi_{v,u}\ket{1}_u\bra{1}_u}$ on every neighbor of $v$. To show this, we use the recursive definition (eq. (\ref{eq:RecursiveDefinitionOfWeightedGraphState}):
\begin{gather}
    X_v \prod_{u\in n(v)} e^{i\chi_{v,u}\ket{1}_u\bra{1}_u} \ket{\psi}=X_v \prod_{u\in n(v)} e^{i\chi_{v,u}\ket{1}_u\bra{1}_u} \frac{1}{\sqrt{2}}\left(\ket{0}_{v}\ket{\phi}+\ket{1}_v\prod_{u\in n(v)}e^{-i\chi_{v,u}\ket{1}_u\bra{1}_u}\ket{\phi}\right)
    \nonumber \\
    =\frac{1}{\sqrt{2}}\left(\ket{0}_{v}\ket{\phi}+\ket{1}_v\prod_{u\in n(v)}e^{i\chi_{v,u}\ket{1}_u\bra{1}_u}\ket{\phi}\right) \ ,
    \label{eq:flipping signs}
\end{gather}
which by the recursive definition (eq. (\ref{eq:RecursiveDefinitionOfWeightedGraphState})) is the same weighted graph states with all the weights $\chi_{v,u}$ flipped. Hence, assuming the weighted graph state is 1-dim, one can apply this flipping operations on $c_2$, then on the qubit to the right of $c_2$, then to the qubit to its right, etc., which will, in the end, correct all the signs so that the original weights are restored, as shown in figure \ref{fig:FlippingSigns}.}

{In appendix \ref{Appendix:ProofOfLogicalQubitWeightedGraphState} we prove that if the original weighted graph state consists only of the three qubits $a,b_1,b_2$ (a scenario in which the Schmidt rank argument does not hold), then in order to get the state (\ref{eq:DesiredWeightedGraphStateWithb1b2BeingLogicalQubit}) from the state (\ref{eq:WeightedGraphStateAfterApplyngGeneralProjectionOnQubita}) using only single-qubit rotations on $b_1$,$b_2$, the two cases we mentioned are still the only possible ones.}

\begin{figure}
    \centering
    \includegraphics[width=0.5\linewidth]{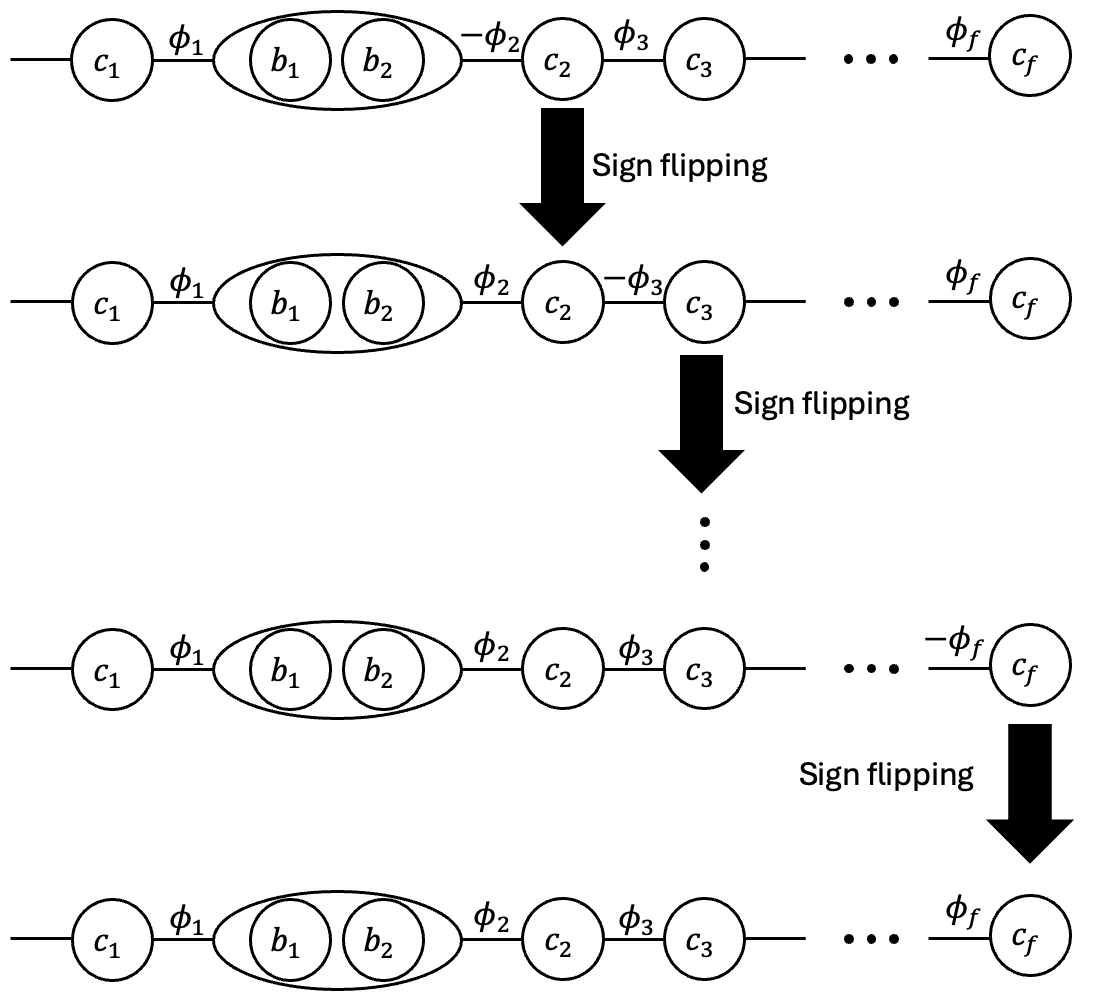}
    \caption{{By applying a chain of flipping signs operations (see eq. (\ref{eq:flipping signs})), one can carry the flipped sign to the right until it is removed.}}
    \label{fig:FlippingSigns}
\end{figure}


Note that this situation is different from the creation of a logical qubit in a graph state in subsection \ref{Subsec:Creating a logical qubit}, because for each of the cases we have only one state of $a$ on which we can project, unlike the graph state case, for which there are two such states (if $\chi=\pi$ then both cases become the same, as $\chi_1=\chi_2=\pi$, as this is the exact case as for graph states except the fact that other weights in the weighted graph can differ from $\pi$). This fact will be important for the failure case of Type-II fusion in subsection \ref{Subsec:Fusion type-II for weighted graph state}.

So, assuming we are not in the case $\chi_1=\chi_2=\pi$, in both cases we need to project onto a specific state --- and unlike the $X$ measurement that creates a logical qubit with probability one, the probability for projecting on the required state is (in both cases):

\begin{gather}
    p=\frac{1-\cos{\chi}}{4} \ ,
    \label{eq:probability of X-like measurement to succeed}
\end{gather}
which is less than $0.5$ as long as $\chi \ne \pi$. In the case of failure, one can apply a $Z_{b_1}Z_{b_2}$ measurement that will leave us with one of the four components of eq. (\ref{eq:WeightedGraphStateAfterApplyngGeneralProjectionOnQubita}) (without the qubits $b_1$ and $b_2$), which are two separated one-dimensional weighted graph states arising from erasing $a,b_1,b_2$ and their edges (up to single-qubit rotations).

From this we can conclude that the only apparent way to make a logical qubit in one-dimensional weighted graph state with certainty is by creating in this one-dimensional weighted graph two consecutive edges with $\pi$-weight, and applying $X$ measurement on the qubit between those edges. 

\if{Also, assuming we are not in the case $\chi_1=\chi_2=\pi$, in both cases we need to project onto a specific state --- we will now compute the probability for each of those projections, using the following claim.

\begin{claim}
    Assume we have a weighted graph state with wave function $\ket{\Phi}$, and a specific qubit $h$ from this weighted graph. Then:
    \begin{gather}
        \bra{\Phi}e^{-i\phi \ket{1}_h \bra{1}_h}\ket{\Phi}=\frac{1+e^{-i\phi}}{2} \ .
    \end{gather}
\end{claim}

ֿ\begin{proof}
    When using the definition (\ref{eq:DefinitionOfWeightedGraphState}) and opening brackets (including $\ket{+}=\frac{\ket{0}+\ket{1}}{\sqrt{2}}$), we get a representation of $\Phi$ in the standard base of $\prod_{i\in V} \{\ket{0}_i,\ket{1}_i\}$, with all the coefficients being of norm $2^{-\frac{|V|}{2}}$ and with different phases. Thus, half of those terms will include $\ket{0}_a$ and will not be affected by operating with $e^{-i\phi \ket{1}_h \bra{1}_h}$, while the over half of the terms will include $\ket{1}_a$ and operating with $e^{-i\phi \ket{1}_h \bra{1}_h}$ will multiply those coefficients by $e^{-i\phi}$. Thus, when computing $\bra{\Phi}e^{-i\phi \ket{1}_h \bra{1}_h}\ket{\Phi}$ in this basis we will get:
    \begin{gather}
        \bra{\Phi}e^{-i\phi \ket{1}_h \bra{1}_h}\ket{\Phi}=2^{N-1}\cdot \left(2^{-\frac{N}{2}}\right)^2+2^{N-1}\cdot \left(2^{-\frac{N}{2}}\right)^2e^{-i\phi}=\frac{1+e^{-i\phi}}{2} \ .
    \end{gather}
\end{proof}



Now, using that, one can compute the probability for projecting on the desired 1-qubit state in both case 1 and case 2. In both cases, this is the norm of (\ref{eq:WeightedGraphStateAfterApplyngGeneralProjectionOnQubita}). For case 1 we get:

\begin{gather}
    p=
\end{gather}}\fi

\subsection{Do Y-like measurements of one-dimensional weighted graph states exist?}
\label{subsec:Y-like Measurement of 1-dim weighted graph state}
\begin{figure}
    \centering
    \includegraphics[width=0.5\linewidth]{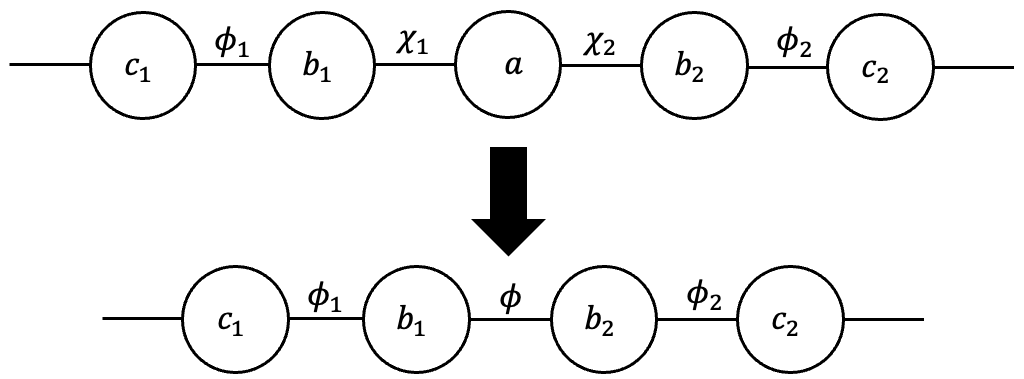}
    \caption{{Y-like measurement on weighted graph states; the goal is to measure the middle qubit $a$, such that the resulting state will be equivalent to a weighted graph state with some new weighted edge between the qubits $b_1,b_2$. We showed that this is not possible if only single-qubit rotations on $b_1,b_2$ are allowed.}}
    \label{fig:YmeasurementWeightedGraphState}
\end{figure}

While X measurements on one-dimensional graph states create logical qubits, Y measurement of a qubit will erase it and its edges, while connecting its neighbors by an edge (up to single-qubit rotations). In this subsection we will discuss when this is possible for weighted graph states --- meaning the resulting wave function (\ref{eq:WeightedGraphStateAfterApplyngGeneralProjectionOnQubita}) is (up to single-qubit rotations on $b_1$ and $b_2$) a new weighted graph state, where instead of $a$ and its edges, there is a new edge between $b_1$ and $b_2$ with weight $\phi$ {(see figure \ref{fig:YmeasurementWeightedGraphState})}:

\begin{gather}
    \ket{0}_{b_1}\ket{0}_{b_2}\ket{\phi_{b_1}}\ket{\phi_{b_2}}+ \nonumber \\
    +\ket{1}_{b_1}\ket{0}_{b_2}e^{-i\phi_{1}\ket{1}_{c_1}\bra{1}_{c_1}}\ket{\phi_{b_1}}\ket{\phi_{b_2}}+\nonumber \\
    +\ket{0}_{b_1}\ket{1}_{b_2}e^{-i\phi_{2}\ket{1}_{c_2}\bra{1}_{c_2}}\ket{\phi_{b_1}}\ket{\phi_{b_2}}+\nonumber \\
    +e^{-i\phi}\ket{1}_{b_1}\ket{1}_{b_2}e^{-i\phi_{1}\ket{1}_{c_1}\bra{1}_{c_1}}e^{-i\phi_{2}\ket{1}_{c_2}\bra{1}_{c_2}}\ket{\phi_{b_1}}\ket{\phi_{b_2}} \ .
    \label{eq:DesiredWeightedGraphStateWithb1b2BeingConnectedByNewEdgeWithWeightPhi}
\end{gather}

In \cite{yamazaki2025measurementbasedquantumcomputationweighted}, the authors introduced a protocol for Y-like measurements on weighted graph states with a uniform weight. In this subsection, we focus on whether this is possible when allowing only single-qubit rotations on $b_1$ and $b_2$, for which our answer is negative.

If the rotation on $b_1$ mixes $\ket{0}_{b_1}$ and $\ket{1}_{b_1}$, then from the first term of eq. (\ref{eq:WeightedGraphStateAfterApplyngGeneralProjectionOnQubita}) we will have $\ket{1}_{b_1}\ket{0}_{b_2}\ket{\phi_{b_1}}\ket{\phi_{b_2}}$ term, which no other term can cancel. Thus, the rotation on $b_1$ needs to be diagonalized in the basis $\ket{0}_{b_1},\ket{1}_{b_1}$. For the same reason, the rotation on $b_2$ needs to be diagonalized in the basis $\ket{0}_{b_2},\ket{1}_{b_2}$. This means that all the four terms in eq. (\ref{eq:WeightedGraphStateAfterApplyngGeneralProjectionOnQubita}) will be multiplied by phases.
In particular, the absolute value of the coefficient of all those terms is preserved, and in eq. (\ref{eq:DesiredWeightedGraphStateWithb1b2BeingConnectedByNewEdgeWithWeightPhi}) those values are all the same, so in order to get the state  (\ref{eq:DesiredWeightedGraphStateWithb1b2BeingConnectedByNewEdgeWithWeightPhi}) from the state (\ref{eq:WeightedGraphStateAfterApplyngGeneralProjectionOnQubita}) by rotations on $b_1,b_2$ one must have the same absolute value for all the coefficients in eq. (\ref{eq:WeightedGraphStateAfterApplyngGeneralProjectionOnQubita}):

\begin{gather}
    |A+B|=|A+Be^{-i\chi_1}|=|A+Be^{-i\chi_2}|=|A+Be^{-i\left(\chi_1+\chi_2\right)}| \ .
\end{gather}

As explained in appendix \ref{Appendix:ProofOfLogicalQubitWeightedGraphState}, we conclude from that:

\begin{gather}
    \arg A - \arg B = \pm \left(\arg A - \arg B + \chi_1 \right) = \pm \left(\arg A - \arg B + \chi_2 \right) =\pm \left(\arg A - \arg B + \chi_1 + \chi_2 \right) \ . 
\end{gather}

Assuming $\chi_1,\chi_2\ne 0$, we must have minus signs before the second and third terms, hence $\chi_1=\chi_2$, and because $\chi_1,\chi_2\ne 0$ we also must have plus sign before the fourth term, yielding $\chi_1+\chi_1=0$, meaning $\chi_1=\chi_2=\pi$, and we are left with the case where the edges between $a$ and $b_1,b_2$ are like those of a regular graph and a regular Y measurement will work.

\subsection{Creating 2-qubit weighted graph states from GHZ states}
\label{Subsec:Creating 2-qubit weighted graph states from GHZ states}

In \cite{PhysRevA.85.012307,PhysRevResearch.5.023124} it was shown that a graph state can be generated by measurements on weighted graph states. In particular, \cite{PhysRevResearch.5.023124} generates a GHZ state by measurements on a weighted graph state. In this subsection we show the opposite --- that a 2-qubit weighted graph state with a target weight can be generated by measurements on a GHZ state with 100 percent probability. Obviously, after a measurement on a GHZ state, the resulting 2-qubit state is equivalent to some 2-qubit weighted graph state \cite{hein2006entanglementgraphstatesapplications}; the important thing here is achieving the desired target weight (instead of achieving a certain weight with some probability and another weight with the complementary probability). Furthermore, we show that there is a range of target weights that can be generated by measurement on three-qubit one-dimensional weighted graph states, with the range depending on the original weights.

\begin{figure}
    \centering
    \includegraphics[width=0.3\linewidth]{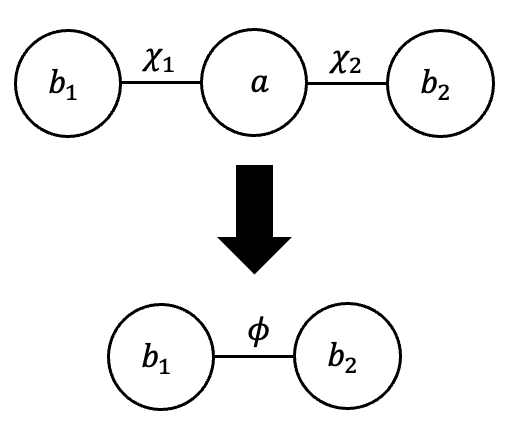}
    \caption{{Given a 1-dim three-qubit weighted graph state with weights $\chi_1,\chi_2$, the goal is to measure the middle qubit $a$ such that in both measurement outcomes, the weight of the resulting two-qubit weighted graph state will be the same value $\phi$. This is possible for any weight in the range (\ref{eq:RangeOfTheNewWeightBetweenB1AndB2AfterYLikeMeasurement}). In the case that at least one of $\chi_1,\chi_2$ is $\pi$, any measurement on the middle qubit will result in the same weight $\phi$ for both measurement outcomes, and the range (\ref{eq:RangeOfTheNewWeightBetweenB1AndB2AfterYLikeMeasurement}) becomes all the possible weights.}}
    \label{fig:3qubitsweightedTo2qubitsweighted}
\end{figure}

Assuming the original weighted graph state consist only of the three qubits $a,b_1,b_2$ {as in figure \ref{fig:3qubitsweightedTo2qubitsweighted}} (again meaning that single-qubit rotations on $b_1,b_2$ are less constrained because we don't have the limitations of $\ket{\phi_{b_1}}$ etc.), we measure some local operator of qubit $a$ with the eigenvectors (up to global phase) $A\ket{0}_a+B\ket{1}_a$ and $B^*\ket{0}_a-A^*\ket{1}_a$. In appendix \ref{Appendix:ProofOfCreatingNewEdgeBetweenB1AndB2} we prove that the condition for the outcome weight to be the same for the two possible measurement outcomes is:

\begin{gather}
\label{eq:ConditionsOnAandBInOrderToCreateNewWeightEdge}
    \arg B = \arg A + \frac{\chi_1+\chi_2\pm \pi}{2} \ ,
\end{gather}
assuming that $-\pi<\chi_1,\chi_2<\pi$, and if one of $\chi_1,\chi_2$ is $\pi$ then there is no condition. Given this condition holds, the resulting weight is:

\begin{gather}
\label{eq:NewWeightBetweenB1AndB2AfterYLikeMeasurement}
    \phi=\pm \arccos{\left(1-2|A|^2|B|^2\left(1-\cos{\chi_1}\right)\left(1-\cos{\chi_2}\right)\right)} \ ,
\end{gather}

where the sign can be flipped if desired. Also by appendix \ref{Appendix:ProofOfCreatingNewEdgeBetweenB1AndB2} this weight can take any value in the range:

\begin{gather}
\label{eq:RangeOfTheNewWeightBetweenB1AndB2AfterYLikeMeasurement}
    0\leq|\phi|\leq\arccos{\left(1-\frac{1}{2}\left(1-\cos{\chi_1}\right)\left(1-\cos{\chi_2}\right)\right)} \ .
\end{gather}

In summary, exactly the weights in the range (\ref{eq:RangeOfTheNewWeightBetweenB1AndB2AfterYLikeMeasurement}) can be achieved with 100 percent probability by measuring the middle qubit in a 3-qubit one-dimensional weighted graph with weights $\chi_1,\chi_2$. In the case of starting with a 3-qubit one-dimensional graph state that is equivalent to a GHZ state, $\chi_1=\chi_2=\pi$ and the range (\ref{eq:RangeOfTheNewWeightBetweenB1AndB2AfterYLikeMeasurement}) becomes all the possible weights with no restrictions. In fact, this is the case also where only one of $\chi_1,\chi_2$ is $\pi$, and as discussed, in this case there is no condition on $A,B$ --- so any local operator of the middle qubit that we will measure will result in the same equivalent weighted graph state for both measurement outcomes.

\if{In appendix \ref{Appendix:ProofOfCreatingNewEdgeBetweenB1AndB2} we prove that if the original weighted graph state consist only from the three qubits $a,b_1,b_2$ (again meaning that single-qubit rotations on $b_1,b_2$ can get us more because we don't have the limitations of $\ket{\phi_{b_1}}$ etc.) then the condition for getting (\ref{eq:DesiredWeightedGraphStateWithb1b2BeingConnectedByNewEdgeWithWeightPhi}) by measuring $a$ in (\ref{eq:WeightedGraphStateAfterApplyngGeneralProjectionOnQubita}) (up to single-qubit rotations) is:

\begin{gather}
\label{eq:ConditionsOnAandBInOrderToCreateNewWeightEdge}
    \arg B = \arg A + \frac{\chi_1+\chi_2\pm \pi}{2} \ ,
\end{gather}
assuming that $-\pi<\chi_1,\chi_2<\pi$, and if one of them is $\pi$ then there is no condition. If this condition is satisfied, then the resulting wave function (\ref{eq:WeightedGraphStateAfterApplyngGeneralProjectionOnQubita}) is indeed the desired new weighted graph state, with the new weight being:

\begin{gather}
\label{eq:NewWeightBetweenB1AndB2AfterYLikeMeasurement}
    \phi=\pm \arccos{\left(1-2|A|^2|B|^2\left(1-\cos{\chi_1}\right)\left(1-\cos{\chi_2}\right)\right)} \ .
\end{gather}
Also by appendix \ref{Appendix:ProofOfCreatingNewEdgeBetweenB1AndB2} this weight can get any value in the range:

\begin{gather}
    0\leq|\phi|\leq\arccos{\left(1-\frac{1}{2}\left(1-\cos{\chi_1}\right)\left(1-\cos{\chi_2}\right)\right)} \ .
\end{gather}

Note that in the case where $\chi_1=\chi_2=\pi$, we have:

\begin{gather}
\label{eq:NewWeightBetweenB1AndB2AfterYLikeMeasurementWhereChi1AndChi2ArePi}
    \phi=\pm \arccos{\left(1-8|A|^2|B|^2\right)} \ ,
\end{gather}
and $\phi$ can get any value. Also, because in this case we have no condition on $\arg A - \arg B$, we can measure an operator whose two orthogonal eigen-states produce the same $\phi$, thus making this process certain instead of probabilistic. That means that if we begin with 3-qubit one-dimensional graph state (i.e GHZ gate), then by measuring the appropriate operator, one can create a weighted pair with a desired weight.}\fi

\if{A similar approach was presented in \cite{PhysRevResearch.5.023124} where the authors showed that a GHZ gate can be generated by measurements on weighted graph states (while we showed that weighted graph states can be generated from measurements on GHZ states).}\fi

\if{Also, because in this case we have no condition on $\arg A - \arg B$, we can measure an operator whose two orthogonal eigen-states produce the same $\phi$, thus making this process certain instead of probabilistic. That means that if we begin with one-dimensional graph state, we can replace any two consecutive edges with a weighted edges, and by that we can build a weighted graph state with the desired weights and half of the edges.}\fi

\subsection{Type-II fusion for weighted graph states}
\label{Subsec:Fusion type-II for weighted graph state}
\begin{figure}
    \centering
    \includegraphics[width=0.5\linewidth]{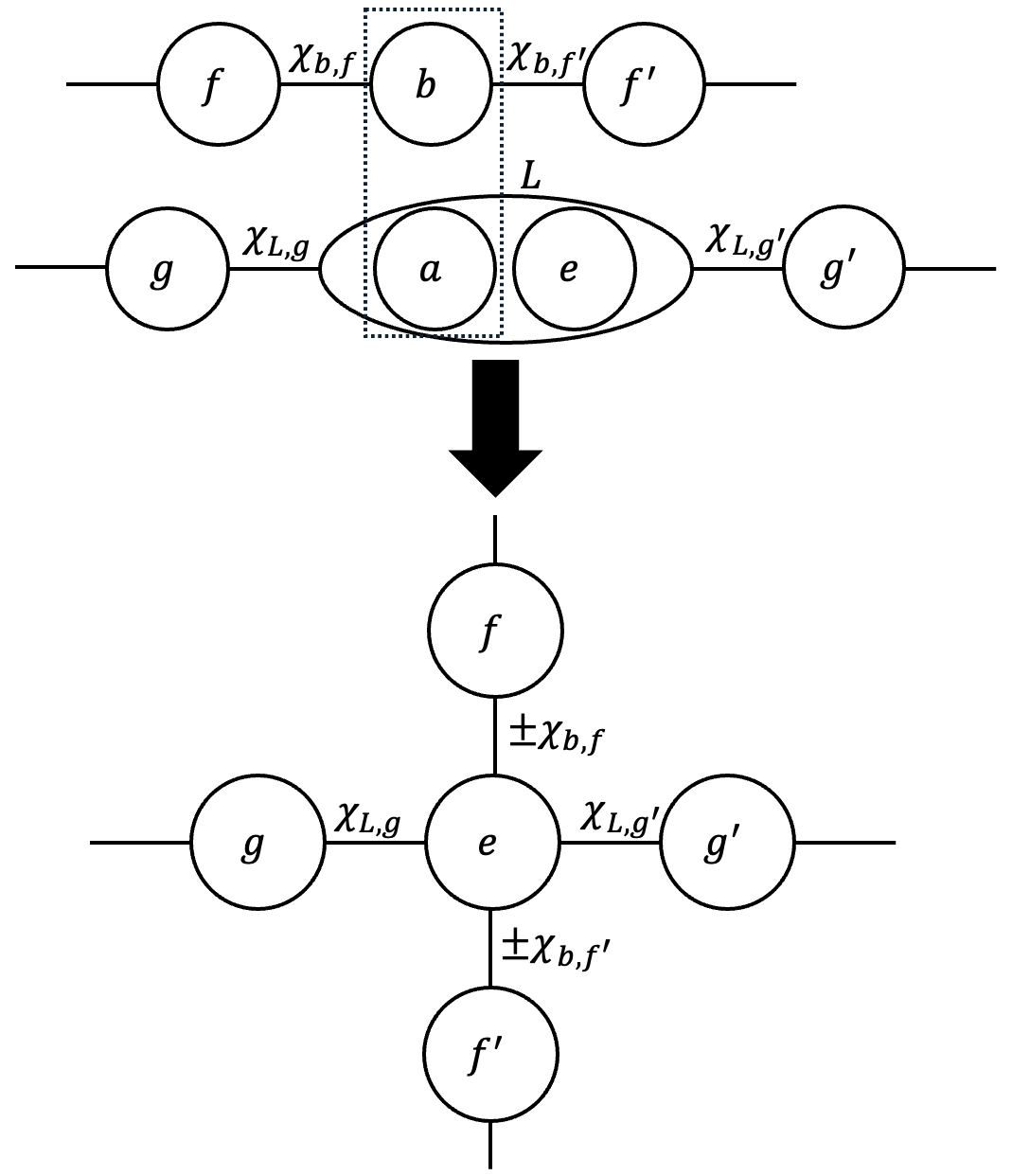}
    \caption{{Type-II fusion of weighted graph states. We begin with two 1-dim weighted graph states, one with a logical qubit that consists of the qubits $a,e$. We then apply the Type-II fusion gate on the qubit $a$ and another qubit $b$ from the other 1-dim weighted graph state, and if the fusion succeeds, the resulting state is a 2-dim weighted graph state in which $e$ inherited the neighbors of $L,b$. The problem occurs when the fusion fails, as unlike the graph states case, this may result in destruction of some of the structure of the original 1-dim weighted graph states.}}
    \label{fig:FusionTypeIIWeightedGraphState}
\end{figure}

Assume that we have two one-dimensional weighted graph states that we wish to fuse. Assume that one of these weighted graph states already contains a logical qubit $L$ consisting of the qubits $a$ and $e$, and choose a qubit $b$ from the other weighted graph{, as illustrated in figure \ref{fig:FusionTypeIIWeightedGraphState}}. By equations (\ref{eq:RecursiveDefinitionOfWeightedGraphState}) and (\ref{eq:LogicalQubit}), the total wave function is:

\begin{gather}
        \ket{\psi}=\left(\ket{0}_{a}\ket{0}_{e}\ket{\phi_L}+\ket{1}_a\ket{1}_{e}\prod_{g\in n(L)}e^{-i\chi_{Lg}\ket{1}_g\bra{1}_g}\ket{\phi_L}\right)\cdot \nonumber \\
    \cdot \left(\ket{0}_{b}\ket{\phi_b}+\ket{1}_b\prod_{f\in n(b)}e^{-i\chi_{bf}\ket{1}_f\bra{1}_f}\ket{\phi_b}\right) \ .
\label{eq:StartingWaveFunctionFusionTypeIIWeightedGraphStates}
\end{gather}

{We now apply the Type-II fusion described in subsection \ref{Subsec:Fusion Type-II}. In case of success, we operate on the state (\ref{eq:StartingWaveFunctionFusionTypeIIWeightedGraphStates}) with one of the operators in eq. (\ref{eq:OperatorFusionTypeIISuccess}), yielding one of the states:

\begin{gather}
    \ket{\psi_{new}}=\frac{1}{\sqrt{2}}\left(\ket{0}_{e}\ket{\phi_L}\ket{\phi_b}+ \ket{1}_e\prod_{g\in n(L)}e^{-i\chi_{Lg}\ket{1}_g\bra{1}_g}\prod_{f\in n(b)}e^{-i\chi_{bf}\ket{1}_f\bra{1}_f}\ket{\phi_L}\ket{\phi_b}\right) \ ,
    \nonumber \\
    \ket{\psi_{new}}=\frac{1}{\sqrt{2}}\left(\ket{0}_{e}\prod_{f\in n(b)}e^{-i\chi_{bf}\ket{1}_f\bra{1}_f}\ket{\phi_L}\ket{\phi_b}+ \ket{1}_e\prod_{g\in n(L)}e^{-i\chi_{Lg}\ket{1}_g\bra{1}_g}\ket{\phi_L}\ket{\phi_b}\right) \ .
\end{gather}
In the second option one can apply the operators multiplications $\prod_{f\in n(b)}e^{i\chi_{bf}\ket{1}_f\bra{1}_f}$ and get the first option with the signs of $\chi_{b,f}$ being flipped, 
which by eq. (\ref{eq:RecursiveDefinitionOfWeightedGraphState}) is the desired fused graph state up to the flipped signs} {as illustrated in figure \ref{fig:FusionTypeIIWeightedGraphState}.} {The flipped signs can be fixed in a similar manner to as illustrated in figure \ref{fig:FlippingSigns}.}

In case of failure, we operate on the state (\ref{eq:StartingWaveFunctionFusionTypeIIWeightedGraphStates}) with one of the operators in eq. (\ref{eq:OperatorFusionTypeIIFailure}). This keeps the two original weighted graph states separated. For the first weighted graph, the situation is the same as for graph states --- "measuring $X_a$" replaces the logical qubit by $e$, up to applying $Z_e$ in the case of a minus, thus leaving us with regular one-dimensional weighted graph state. But, for the second weighted graph, "measuring $X_b$" is problematic; a projection onto $\frac{1}{\sqrt{2}}\left(\ket{0}_b-\ket{1}_b\right)$ will create a logical qubit if we assume case 2 from subsection \ref{Subsec:Creating a logical qubit for 1-dim weighted graph state}, and projection onto $\frac{1}{\sqrt{2}}\left(\ket{0}_b+\ket{1}_b\right)$ will spoil the weighted graph state. This can be fixed as in subsection \ref{Subsec:Creating a logical qubit for 1-dim weighted graph state} by measuring $Z$ operators for the neighbors of $b$, but that will split this one-dimensional weighted graph state into two smaller ones. The effect is similsar to attempting Type-I fusion to fuse two graph states using middle qubits --- which is one of the main reasons why Type-II fusion is used in this case rather than Type-I fusion (as for graph state it yields "good" state upon failure), thus eliminating the advantage of this fusion process over the Type-I.
In subsection \ref{subsec:Generalized fusion type-II for weighted graph state} we show that choosing another Type-II fusion process as in subsection \ref{subsec:Generalized Type-II Fusion} cannot solve this problem --- as long as there is a success probability for the fusion process, the possible resulting states that are product states will not all result from the same projection onto a $b$ state (even if we ignore the fact that the projection onto $a$ is also under restriction for the non-success case to work as in the regular Type-II fusion case). 

\begin{figure}
    \centering
    \includegraphics[width=0.5\linewidth]{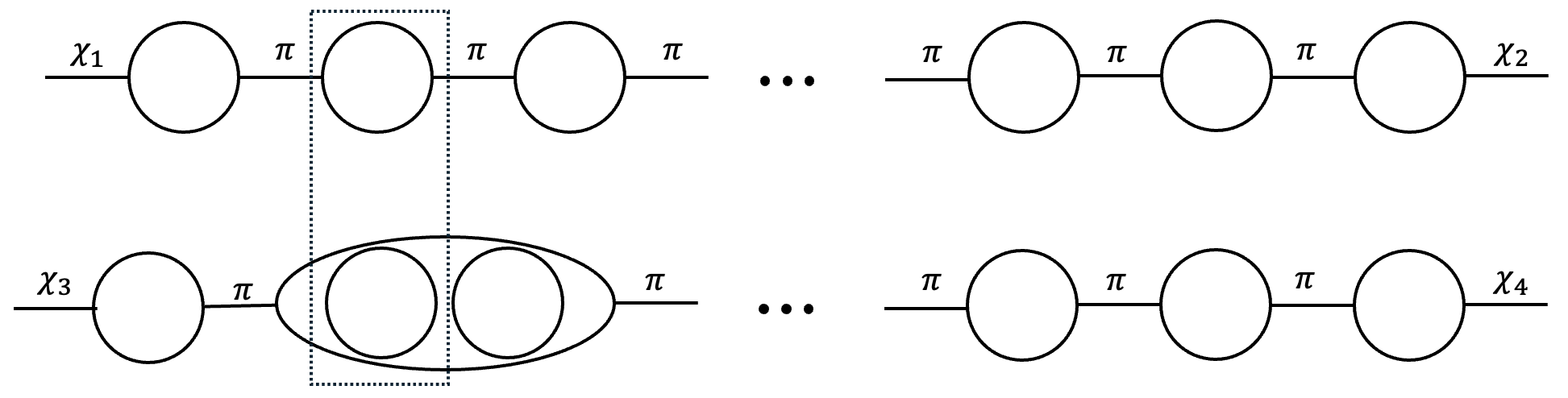}
    \caption{{Padding the original 1-dim weighted graph states with a chain of $\pi$-weights, in order to apply regular Type-II fusion until success --- thus preventing a destruction of the structure in the case of fusion failure.}}
    \label{fig:AddingManyPiWeights}
\end{figure}

{One can overcome this problem by constructing in advance many $\pi$-weights between the other weights in the 1-dim weighted graph states, as illustrated in figure \ref{fig:AddingManyPiWeights}. In this way, a regular Type-II fusion can be made between qubits that have two $\pi$ edges, until the fusion succeeds, and then $X,Y$ measurements can erase the remaining $\pi$-weights and their qubits (see appendix \ref{appendix:Erasing added pi weights chain} for full explanation about the erasing process). This solution requires many additional qubits and Type-I fusion operations, such that for any Type-II fusion the probability to fail repeatedly until the relevant $\pi$-weights are exhausted is negligible. If we have, for example, $n_1$ Type-II fusion operations needed, and $n_2$ additional $\pi$-weights for every 1-dim weighted graph state per Type-II fusion process in which it is involved, then the probability that all the Type-II fusions needed will not run out of $\pi$-weights is:
\begin{gather}
    p=\left(1-0.5^{n_2+1}\right)^{n_1} \ ,
\end{gather}
(notice we have $0.5$ success probability for the weighted graph states themselves even if all the $n_2$ $\pi$-weights do run out). Thus, if we want this probability to be greater than some $1-\epsilon$, it requires 
\begin{gather}
    n_2>\log_2{\frac{1}{2\left(1-\left(1-\epsilon\right)^{\frac{1}{n_1}}\right)}}=\log_2 n_1+\log_2 \frac{1}{\epsilon}-1+O\left(\frac{1}{n_1}\right) \ .
\end{gather}
The number of additional needed qubits and Type-I fusion operations will be 
\begin{gather}
    n_{tot}=n_1\left(2n_2+2\right)=2n_1\log_2 n_1+2\left(\log_2\left(\frac{1}{\epsilon}\right)-1\right)n_1+O\left(1\right) \ ,
    \label{eq:ntot}
\end{gather}
which will not prevent scaling.} 

It is also important to note that the probabilities of success remain the same, for the same reason as in subsection \ref{Subsec:Fusion type-I for weighted graph state}. We now denote $f_1=\ket{0}_{e}\ket{\phi_L}$,$f_2=\ket{1}_{e}\prod_{g\in n(L)}e^{-i\chi_{Lg}\ket{1}_g\bra{1}_g}\ket{\phi_L}$ and $f_3=\ket{0}_{b}$,$f_4=\prod_{f\in n(b)}e^{-i\chi_{bf}\ket{1}_f\bra{1}_f}\ket{\phi_b}$, so the initial wave function (\ref{eq:StartingWaveFunctionFusionTypeIIWeightedGraphStates}) takes the form (\ref{eq:StartingWaveFunctionFusionTypeIGraphStatesInf1f2f3f4Language}). While the couple $f_3,f_4$ is not orthonormal, the couple $f_1,f_2$ is (because $f_1,f_2$ contain $\ket{0}_e$,$\ket{1}_e$ respectively), 
{\color{black}and as explained in subsection \ref{Subsec:Fusion Type-II}, this does not affect the computation of the probabilities of the possible resulting states in the case of success. In the case of failure, the total probability is still half, which is now divided between a projection onto $(f_1+f_2)(f_3-f_4)$ and onto $(f_1-f_2)(f_3+f_4)$. Using eq. (\ref{eq:The inner product of f3 and f4 for weighted graph}) for the value of $z=f_4^\dag f_3$, the probabilities of failure have ratio $|f_3-f_4|^2:|f_3+f_4|^2=\left(2-2\Re{f_4^\dag f_3}\right):\left(2+2\Re{f_4^\dag f_3}\right)$. Hence, the failure probabilities read:

\begin{gather}
    p_{\left(\bra{0}_a\pm \bra{1}_a\right)\left(\bra{0}_b\mp \bra{1}_b\right)}=\frac{1}{4}\left(1\mp Re(z)\right) \ ,
\end{gather}

where (using eq. (\ref{eq:The inner product of f3 and f4 for weighted graph})):

\begin{gather}
    Re(z)=\frac{1+\cos{\chi_{b,f}}+\cos{\chi_{b,f'}}+\cos{\left(\chi_{b,f}+\chi_{b,f'}\right)}}{4} \ .
\end{gather}

Note that $\chi_{b,f},\chi_{b,f'}$ are the weights on the edges containing the qubit $b$ on its original weighted graph state. Specifically, if $\chi_{b,f'}=-\chi_{b,f}=\chi$ (case 2 from subsection \ref{Subsec:Creating a logical qubit for 1-dim weighted graph state}), then $\Re{z}=\frac{1+\cos{\chi}}{2}$, and the good failure that occurs for the projection $\left(\bra{0}_a+ \bra{1}_a\right)\left(\bra{0}_b- \bra{1}_b\right)$ has probability:

\begin{gather}
    p_{\left(\bra{0}_a+ \bra{1}_a\right)\left(\bra{0}_b- \bra{1}_b\right)}=\frac{1-\cos{\chi}}{8} \ ,
\end{gather}

which is less than $\frac{1}{4}$ assuming $\chi\ne \pi$ (and equality is achieved when $\chi=\pi$, which is the case of regular graph fusion). Thus, not only can the failure case of the Type-II fusion be problematic, but the problematic failure probability is also larger than the good failure probability.

Finally, by the same arguments, any fusion process which results in Bell projection will retain its success probability (for example, Bell projection with ancillary qubits that boost its probability of success). That is because the projected expressions will retain their norms (from the exact reason explained here). In the next subsection, we show that this still holds for any fusion involving projection onto maximally entangled state of the qubits $a,b$ (not necessarily Bell projection). But this does not hold for fusion processes utilizing non-maximally entangled projections, which we will discuss in the next subsection. {It is also worth noting that using ancilla qubits to boost the success probability can be relevant for the addition of $\pi$-weights option, as it will replace the base $2$ in (\ref{eq:ntot}) with base $\frac{1}{1-p}$ where $p$ is the probability of success of the fusion, thus lowering the number of additional qubits and Type-I fusion overall (the number of added ancilla qubits will be linear in $n_1$ while we will decrease the coefficient of $n_1\ln n_1$).}

\if{\begin{gather}
    p_{\left(\bra{0}_a+ \bra{1}_a\right)\left(\bra{0}_b- \bra{1}_b\right)}=\frac{3-2\cos{\chi}-\cos{2\chi}}{16}=\frac{2-\cos{\chi}-\cos^2{\chi}}{4} \ .
\end{gather}

This is a quadratic function of $\cos{\chi}$ with maximum $\frac{9}{16}$ when $\cos{\chi}=\frac{1}{2}$ (so $\chi=\pm \frac{\pi}{3}$).}\fi

}

\if{and as explained in subsection \ref{Subsec:Fusion type-I for weighted graph state}, this does not affect the computation of the probabilities of the possible resulting states. }\fi

\subsection{Generalized Type-II fusion for weighted graph states}
\label{subsec:Generalized fusion type-II for weighted graph state}
\begin{figure}
    \centering
    \includegraphics[width=0.5\linewidth]{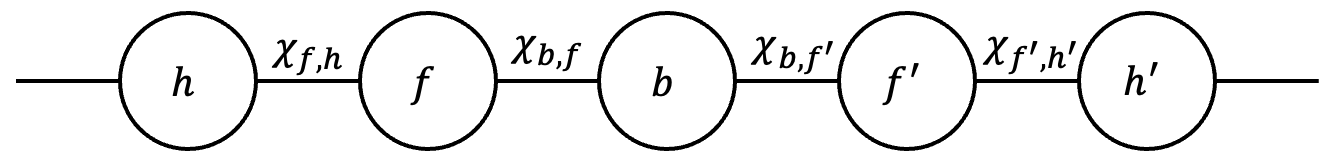}
    \caption{{The 1-dim weighted graph state that contains the qubit $b$. The wave functions $f_3,f_4$ are not orthogonal (unlike the graph states case), and their inner product is given by $\chi_{b,f},\chi_{b,f'}$ as in equation (\ref{eq:The inner product of f3 and f4 for weighted graph}).}}
    \label{fig:bff'hh'}
\end{figure}

{If we apply a general matrix $U$ as in equations (\ref{eq:Unitary transformation}) or (\ref{eq:Unitary transformation with ancilla}) in subsection \ref{subsec:Generalized Type-II Fusion}, then one needs to update the computation of the probabilities and entanglement entropies. If we denote by $f,f'$ the neighbors of $b$ in its original weighted graph, and by $h,h'$ their other original neighbors, respectively (so their order is $h,f,b,f',h'$ {as illustrated in figure \ref{fig:bff'hh'}}), then one can write the wave functions $f_3,f_4$ as (it is the same situation as in eq. (\ref{eq:WritingTheWaveFunctionByDelitingab1b2})):}

\begin{gather}
    f_3=\frac{1}{2}\left(\ket{0}_{f}\ket{\phi_{f}}+\ket{1}_{f} e^{-i\chi_{f,h}\ket{1}_{h}\bra{1}_{h}}\ket{\phi_{f}}\right)\left(\ket{0}_{f'}\ket{\phi_{f'}}+\ket{1}_{f'} e^{-i\chi_{f',h'}\ket{1}_{h'}\bra{1}_{h'}}\ket{\phi_{f'}}\right) \ , \nonumber \\
    f_4=\frac{1}{2}\left(\ket{0}_{f}\ket{\phi_{f}}+e^{-i\chi_{b,f}}\ket{1}_{f} e^{-i\chi_{f,h}\ket{1}_{h}\bra{1}_{h}}\ket{\phi_{f}}\right)\cdot\nonumber \\
    \cdot\left(\ket{0}_{f'}\ket{\phi_{f'}}+e^{-i\chi_{b,f'}}\ket{1}_{f'} e^{-i\chi_{f',h'}\ket{1}_{h'}\bra{1}_{h'}}\ket{\phi_{f'}}\right) \ .
\end{gather}
where $\ket{\phi_{f}},\ket{\phi_{f'}},e^{-i\chi_{f,h}\ket{1}_{h}\bra{1}_{h}}\ket{\phi_{f}},e^{-i\chi_{f',h'}\ket{1}_{h'}\bra{1}_{h'}}\ket{\phi_{f'}}$ are all of norm one, thus the inner product of $f_3,f_4$ is:

\begin{gather}
    z:=f_4^{\dag}f_3=\frac{\left(1+e^{i\chi_{b,f}}\right)\left(1+e^{i\chi_{b,f'}}\right)}{4} \ , \hspace{0.2cm} |z|^2=\frac{\left(1+\cos{\chi_{b,f}}\right)\left(1+\cos{\chi_{b,f'}}\right)}{4} \ .
    \label{eq:The inner product of f3 and f4 for weighted graph}
\end{gather}

Notice that if $b$ has only one neighbor on its original graph, $f$, then one can just set $\chi_{b,f'}=0$.

{If we measure the two photons in the same channel, the resulting state (\ref{eq:Wave function of non-relevant state}) is still a product state and hence we will still call those $(i,i)$ states --- non-relevant states, and the other $(i,j)$ states --- relevant states. The non-relevant state $(i,i)$ will still be the state (\ref{eq:Wave function of non-relevant state}), but the probability and the normalization factor need to be updated using eq. (\ref{eq:The inner product of f3 and f4 for weighted graph}):

\begin{gather}
    p_{ii}=N_{ii}^2=\left(|U_{1i}|^2+|U_{2i}|^2\right)\left(|U_{3i}|^2+|U_{4i}|^2+2\Re{\left(zU_{3i}U_{4i}^*\right)}\right) \ .
    \label{eq:Probability of non-relevant state for weighted graph states}
\end{gather}

The relevant state will still be described by eq. (\ref{eq:Wave function of relevant state}), with the coefficients $a_{ij},b_{ij},c_{ij},d_{ij}$ still being computed by eq. (\ref{eq:The coefficients of the wave function}), but the probability and normalization factor are not the same as in eq. (\ref{eq:The coefficients of the wave function}) but now read as:

\begin{gather}
    N_{ij}=\sqrt{4p_{ij}}=\sqrt{|a_{ij}|^2+|b_{ij}|^2+2\Re{\left(za_{ij}b_{ij}^*\right)}+|c_{ij}|^2+|d_{ij}|^2+2\Re{\left(zc_{ij}d_{ij}^*\right)}} \ .
    \label{eq:normalization factor for relevant state of generalized fusion of weighted graph states}
\end{gather}

If the 2-qubit state we project onto is maximally entangled then $M_{ij}$ in eq. (\ref{eq:Wave function of relevant state}) is unitary multiplied by $\frac{1}{\sqrt{2}}$, hence $a_{ij}b_{ij}^*+c_{ij}d_{ij}^*=0$, and the probability (and normalization factor) $p_{ij}$ (\ref{eq:normalization factor for relevant state of generalized fusion of weighted graph states}) is equal to the case of graph state fusion (\ref{eq:The coefficients of the wave function}). Thus, any fusion process that involves projection onto a 2-qubit maximally entangled state, will retain its success probability (because the probability of every relevant state will remain the same). However, unlike Bell projections that also fuse the weighted graph states, a generalized maximally entangled projection that works well for graph states may not behave the same for weighted graph states (see next subsection). This still holds even if the fusion process is not of the form described in subsection \ref{subsec:Generalized Type-II Fusion} --- for example, the fusion process described in \cite{Bartolucci:2021pty}. In those cases instead of $N_{ij}=\sqrt{4p_{ij}}$ there will be other relations, but the relation will not depend on the states being fused and will also remain the same, and the fact that $f_3,f_4$ are not orthogonal will not affect the norm of the projected expression. By \cite{rimock2024generalizedtypeiifusion}, if there are no ancilla qubits and no vacuum ancilla qubits, meaning we use generalized fusion with $U$ being $4$ by $4$ (\ref{eq:Unitary transformation}), the maximum success probability to project onto a 2-qubit maximally entangled state is half, which is achieved when $|U_{1i}|^2+|U_{2i}|^2=\frac{1}{2}$ for every $1\leq i \leq 4$. We can furthermore deduce the maximal probability to project onto a 2-qubit maximally entangled state using ancilla qubits, based on the simulations in \cite{Schmidt:2024rqz}.

As noted, the entanglement entropy of the resulting state (\ref{eq:Wave function of relevant state}) under dividing our Hilbert space into the Hilbert spaces of each of the original weighted graph states, is not the entanglement entropy of the 2-qubit $a,b$ state on which we project, because the pair $f_3,f_4$ is not an orthonormal pair. 
If we denote by $f_5=\frac{f_4-z^*f_3}{\sqrt{1-|z|^2}}$ the wave function orthonormal to $f_3$ such that $Span\{f_3,f_4\}=Span\{f_3,f_5\}$, we can write the wave function (\ref{eq:Wave function of relevant state}) as:

\begin{gather}
    \ket{\phi}_{ij}=
    \begin{bmatrix}
        f_1 & f_2
    \end{bmatrix}
    M_{ij}
    \begin{bmatrix}
        f_3 \\
        f_4
    \end{bmatrix}
    =\begin{bmatrix}
        f_1 & f_2
    \end{bmatrix}
    M_{ij}
    \begin{bmatrix}
        1 & 0 \\
        z^* & \sqrt{1-|z|^2}
    \end{bmatrix}
    \begin{bmatrix}
        f_3 \\
        f_5
    \end{bmatrix} \ ,
\end{gather}
hence we can replace the matrix $M_{ij}$ in equations (\ref{eq:Reduced Density matrix Graph state}),(\ref{eq:Determinant identities Graph states}) with the new matrix:

\begin{gather}
    M'_{ij}=M_{ij}
    \begin{bmatrix}
        1 & 0 \\
        z^* & \sqrt{1-|z|^2}
    \end{bmatrix}
    =\frac{1}{N_{ij}}\begin{bmatrix}
        a_{ij}+z^*b_{ij} & \sqrt{1-|z|^2}b_{ij} \\
        c_{ij}+z^*d_{ij} & \sqrt{1-|z|^2}d_{ij}
    \end{bmatrix}\ .
    \label{eq:Mij prime}
\end{gather}
Thus, the determinant of $\rho_{ij}$ will be:

\begin{gather}
    det(\rho_{ij})=\left(det\left(M'_{ij}\right)\right)^2=\left(det\left(\begin{bmatrix}
        1 & 0 \\
        z^* & \sqrt{1-|z|^2}
    \end{bmatrix}\right)\right)^2\left(det\left(M_{ij}\right)\right)^2\nonumber\\=\left(1-|z|^2\right)\frac{|a_{ij}d_{ij}-b_{ij}c_{ij}|^2}{N_{ij}^4}=\left(1-|z|^2\right)\left|\frac{(U_{1i}U_{2j}-U_{1j}U_{2i})(U_{3j}U_{4i}-U_{3i}U_{4j})}{4p_{ij}}\right|^2 \ .
    \label{eq:Determinant identities Weighted Graph states}
\end{gather}
}
{\color{black}Hence, the determinant (that is monotonic in the entanglement entropy) is multiplied by $1-|z|^2$, while the normalization factor in the denominator is changed by eq. (\ref{eq:normalization factor for relevant state of generalized fusion of weighted graph states}), compared to the determinant accompanying the two-qubit state on which we project. As explained, if the two-qubit state on which we project is maximally entangled, 
then the normalization factor (\ref{eq:normalization factor for relevant state of generalized fusion of weighted graph states}) is the same as for the two-qubit state on which we project (\ref{eq:The coefficients of the wave function}). Thus, in this case the determinant associated with the two-qubit state on which we project, is just multiplied by $1-|z|^2$, and the determinant of the resulting state is $\frac{1-|z|^2}{4}$. Furthermore, if the fusion matrix $U$ is $4$ by $4$ and satisfying $|U_{1i}|^2+|U_{2i}|^2=\frac{1}{2}$ for every $1\leq i \leq 4$, then all the relevant states will have maximal entanglement entropy for the two-qubit state on which we project and total probability half \cite{rimock2024generalizedtypeiifusion}, hence those kinds of matrices will have $\frac{1-|z|^2}{4}$ determinant for all the relevant states with total probability half. If we wish to perform a fusion operation that results in the desired fusion of the two original weighted graphs, and after the projection we allow only single-qubit rotations, then we must project onto a state with determinant $\frac{1-|z|^2}{4}$ as of the desired fused weighted graph state (as single-qubit rotations do not change entanglement entropy, and this is the entanglement entropy that we will get from Bell projections which results in the desired fusion). 
While the half success probability of the fusion cannot be breached by projecting onto 2-qubit maximally entangled states (without using ancilla qubits) \cite{rimock2024generalizedtypeiifusion}, it may be possible to surpass it when some of the relevant states will be the result of a 2-qubit non-maximally entangled projection, while the entanglement entropy of the relevant states itself will be $\frac{1-|z|^2}{4}$ (and furthermore those states will be the desired fused state up to single-qubit rotations). 
In the next subsection, we show (in the case where the qubit $b$ has only one neighbor in its original weighted graph) the existence of a 2-qubit non-maximally entangled state projection that results in the desired fused state up to single-qubit rotations.

\if{But, a maximal entanglement entropy can be achieved --- the condition for that is $M_{ij}'$ being unitary up to multiplication by scalar, meaning:
\begin{gather}
    |a_{ij}+z^*b_{ij}|=\sqrt{1-|z|^2}|d_{ij}| \ , \nonumber \\
    |c_{ij}+z^*d_{ij}|=\sqrt{1-|z|^2}|b_{ij}| \ , \nonumber \\
    a_{ij}b_{ij}^*+c_{ij}d_{ij}^*+z^*\left(|b_{ij}|^2+|d_{ij}|^2\right)=0 \ .
\end{gather}}\fi

\if{For a given value of $z$, such $a_{ij},b_{ij},c_{ij},d_{ij}$ can be the result of a generalized fusion. Define the following unitary $U$:

\begin{gather}
    U=\begin{bmatrix}
        \sqrt{\frac{1}{2}-\frac{1}{2|z|}} & \sqrt{\frac{1}{2}-\frac{1}{2|z|}} & * & * \\
        \sqrt{\frac{2}{|z|}-\frac{2}{|z|^2}} & 0 & * & * \\
        \frac{\sqrt{2}}{|z|}-\frac{1}{\sqrt{2}} & \frac{1}{\sqrt{2}} & * & *\\
        \frac{1}{\sqrt{2}z^*} & -\frac{1}{\sqrt{2}z^*} & * & * 
    \end{bmatrix} \ ,
\end{gather}

where the $*$ are some values that completing $U$ to a unitary (which is possible because the first two columns are orthogonal and from norm one).}\fi 

} 

One can ask whether there exists a fusion protocol in which all the non-successful fusion events are resulting in the same projection operator onto the original one-dimensional weighted graph state that contained $b$. In particular, this means that every non-relevant state with probability greater than zero must result in the same said projection, and by eq. (\ref{eq:Wave function of non-relevant state}) this means that all those $(i,i)$ states must have the same elements $U_{3i},U_{4i}$ up to global phases. But, this means that for each $i,j$ indices for which the non-relevant states $(i,i)$ and $(j,j)$ have non zero probability, the state $(i,j)$ is a product state (because its determinant is zero by eq. (\ref{eq:Determinant identities Weighted Graph states})). Also, for every $i$ for which the probability for the non-relevant state $(i,i)$ is zero, we will get by eq. (\ref{eq:Probability of non-relevant state for weighted graph states}) that either $U_{1i}=U_{2i}=0$ or $U_{3i}=U_{4i}=0$, and from eq. (\ref{eq:Determinant identities Weighted Graph states}) we get that for every other index $j$ the relevant state $(i,j)$ has zero determinant and is a product state. Hence, we see that all the relevant states are product states, meaning that this kind of fusion with "good" non-successful fusion events is impossible if we want the success scenario to have a nonzero probability.

\subsection{Projecting onto a non-Bell state for weighted graph states}

In this subsection we will make the analogous classification of the resulting state of Type-II fusion process with the general projection as described in subsection \ref{Subsec:Projecting onto non-Bell state}, where now the fusion is on two original one-dimensional weighted graph states.
If we apply the same general 2-qubit projection operator as in eq. (\ref{eq:projectionOperatorForGeneralU}) on the state (\ref{eq:StartingWaveFunctionFusionTypeIIWeightedGraphStates}), then the resulting state is (in analogy to eq. (\ref{eq:Cluster state after measuring a,b for general unitary transformation})):

\begin{gather}
    A\ket{0}_e\ket{\phi_L}
    \ket{\phi_b}
    +B\ket{0}_e\ket{\phi_L}\prod_{f\in n(b)}e^{-i\chi_{bf}\ket{1}_f\bra{1}_f}\ket{\phi_b}
    +C\ket{1}_e\prod_{g\in n(L)}e^{-i\chi_{Lg}\ket{1}_g\bra{1}_g} \ket{\phi_L} \ket{\phi_b}
    \nonumber\\
    +D\ket{1}_e\prod_{g\in n(L)}e^{-i\chi_{Lg}\ket{1}_g\bra{1}_g} \ket{\phi_L}\prod_{f\in n(b)}e^{-i\chi_{bf}\ket{1}_f\bra{1}_f} \ket{\phi_b} \ .
    \label{eq:Resulting state after measuring a,b for general unitary transformation Weighted Case}
\end{gather}

{\color{black}Its entanglement entropy upon dividing the Hilbert space of the fused state into the Hilbert spaces of the original weighted graphs minus the measured qubits, can still be computed using equations (\ref{eq:Determinant identities Weighted Graph states}) and (\ref{eq:Entropy and eigenvalues of the reduced density matrix Graph states}), while replacing $a_{ij},b_{ij},c_{ij},d_{ij}$ by $A,B,C,D$, and the normalization factor is given by eq. (\ref{eq:normalization factor for relevant state of generalized fusion of weighted graph states}).
\if{meaning:

\begin{gather}
    \det{\left(\rho\right)}=\left(1-|z|^2\right) \frac{|AD-BC|^2}{\left(|A|^2+|B|^2+|C|^2+|D|^2+2\Re{\left(z\left(AB^*+CD^*\right)\right)}\right)} \ ,
\end{gather}

and the entanglement entropy reads by (\ref{eq:Entropy and eigenvalues of the reduced density matrix Graph states}).}\fi
}

First, if $B=C=0$ and $|A|=|D|$, then by $Z$-rotation on $e$ we can make $A=D$ and get the desired fused weighted graph state, such as in the graph state case in subsection \ref{Subsec:Projecting onto non-Bell state}. {\color{black}Notice that this is also a 2-qubit maximally entangled state projection. The fact that the set of $A,B,C,D$ is the same as for the graph state case, guarantees that that every fusion process of graph states involving those projections, will not only retain its success probability for weighted graph states, but also "will really succeed" --- the output states will still be the desired fused weighted graph state up to single qubit rotations.}

Now, assume that $|n(b)|=1$ from here on, and denote $n(b)=\{f\}$. We want the resulting state to be a weighted graph state. We can build the resulting stateas follows; first, put all the qubits in the $\ket{+}$ state. Second, apply all the $e^{-i\chi_{xy}\ket{1}_x\ket{1}_y\bra{1}_x}\bra{1}_y$ between any two qubits that are connected by a weighted edge, except for the edge we wish to create between $e$ and $f$. Finally, in order to obtain our final state (\ref{eq:Resulting state after measuring a,b for general unitary transformation Weighted Case}) such that it is a weighted graph state up to a single-qubit rotation, we need to operate on $e,f$ with an operator that commutes with all the other operators we already applied, forcing it to be diagonal in the basis $\ket{0}_e\ket{0}_f,\ket{0}_e\ket{1}_f,\ket{1}_e\ket{0}_f,\ket{1}_e\ket{1}_f$. The only possible operator is (up to normalization):

\begin{figure}
    \centering
    \includegraphics[width=0.5\linewidth]{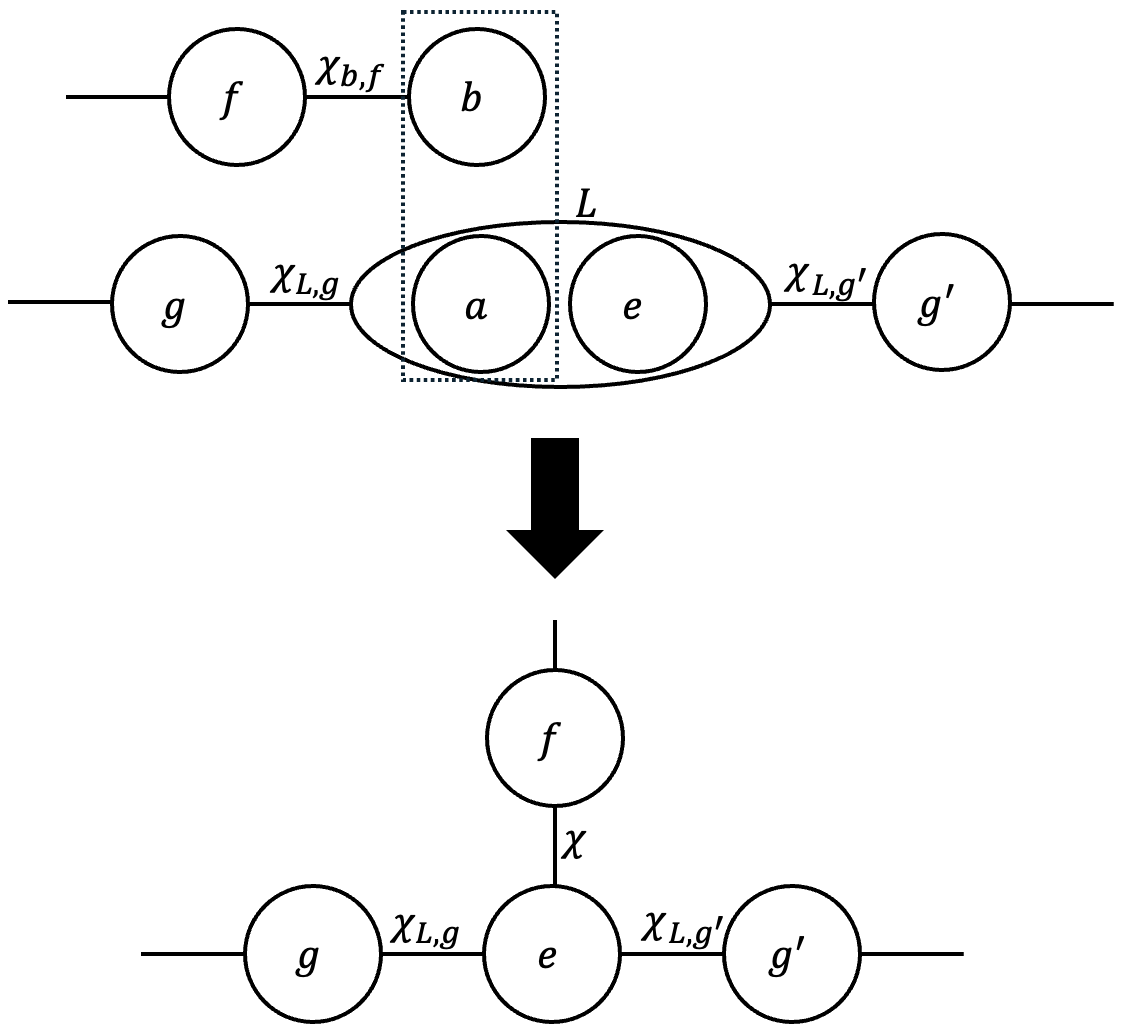}
    \caption{{In the case that $b$ is at the endpoint of its original 1-dim weighted graph state, a non-Bell projection on $a,b$ can result in creating the desired fused weighted graph state, but with the weight $\chi_{b,f}$ changing to a new weight $\chi$ that is given by eq. (\ref{eq:chi by A B C D chibf}).}}
    \label{fig:GeneralizedFusionTypeIInbIsOne}
\end{figure}

\if{Finally, in order to get our final state (\ref{eq:Resulting state after measuring a,b for general unitary transformation Weighted Case}), we need to operate with the operator (up to normalization):}\fi

\begin{gather}
        T_{e,f}=A\ket{0}_e\bra{0}_e+B\ket{0}_e\bra{0}_e e^{-i\chi_{b,f}\ket{1}_f\bra{1}_f}
        \nonumber \\+C\ket{1}_e\bra{1}_e+D\ket{1}_e\bra{1}_e e^{-i\chi_{b,f}\ket{1}_f\bra{1}_f} \ ,
        \label{eq:The 2-qubits gate between e and n(b) for graph state}
\end{gather}

\if{which commute with all the other operators we already applied, making our resulting state a weighted graph state up to single-qubit rotations, }\fi

under the condition that $T_{e,f}$ is a valid operator to apply --- i.e, unitary. In appendix \ref{appendix:Proof of creating weighted graph state by non-Bell projection} we prove that this is equivalent to the equations;

\begin{gather}
    \arg B - \arg A = \arg D - \arg C \in \{\frac{\chi_{b,f}}{2},\frac{\chi_{b,f}}{2}+\pi\} \ ,
    \label{eq:first equation for projecting onto weighted graph with non-Bell state projection}
\end{gather}
and

\begin{gather}
    |A|^2+|B|^2\pm 2|A||B|\cos{\frac{\chi_{b,f}}{2}}=|C|^2+|D|^2\pm 2|C||D|\cos{\frac{\chi_{b,f}}{2}} \ ,
    \label{eq:second equation for projecting onto weighted graph with non-Bell state projection}
\end{gather}
where in eq. (\ref{eq:second equation for projecting onto weighted graph with non-Bell state projection}) the sign in the left side is plus if $\arg B - \arg A = \frac{\chi_{b,f}}{2} $, and minus else, and the same holds for the right side.

One can also compute the resulting weight between $e$ and $f$ using eq. (\ref{eq:State ABCD in basis 00 01 10 11}), which is (see proof A3 in the proof appendix of \cite{rimock2024generalizedtypeiifusion}):

\begin{gather}
    \chi=\arg \left(A+B\right) + \arg \left(C+De^{-i\chi_{b,f}}\right)-\arg \left(A+Be^{-i\chi_{b,f}}\right)-\arg \left(C+D\right) \ .
    \label{eq:chi by A B C D chibf}
\end{gather}

{To summarize, this kind of non-Bell projection, which fuses the original 1-dim weighted graph states, will also change the original weight $\chi_{b,f}$ to the new weight $\chi$ that is given by eq. (\ref{eq:chi by A B C D chibf}), as illustrated in figure \ref{fig:GeneralizedFusionTypeIInbIsOne}}.

{\color{black}An example set of solutions to equations (\ref{eq:first equation for projecting onto weighted graph with non-Bell state projection}) and (\ref{eq:second equation for projecting onto weighted graph with non-Bell state projection}) can be constructed as follows: set $|A|=|D|\ne0,|B|=|C|\ne0$ and $\arg B - \arg A = \arg D - \arg C \in \{\frac{\chi_{b,f}}{2},\frac{\chi_{b,f}}{2}+\pi\}$. Notice that in the case where $|A|=|D|=0$ or $|B|=|C|=0$ we are back to the described final states that are the required fused weighted graph state up to $Z$ rotation on the qubit $e$. Denote $\xi=\pm\frac{|B|}{|A|}=\pm\frac{|C|}{|D|}$, where the sign is plus when $\arg B - \arg A = \arg D - \arg C = \frac{\chi_{b,f}}{2}$, and minus otherwise. Thus, $B=\xi e^{i\frac{\chi_{b,f}}{2}}A$ and $D=\frac{1}{\xi}e^{i\frac{\chi_{b,f}}{2}}C$. Substituting this into eq. (\ref{eq:chi by A B C D chibf}) gives:

\begin{gather}
    \chi=\arg \left((1+\xi e^{i\frac{\chi_{b,f}}{2}})A\right) + \arg \left((1+\frac{1}{\xi} e^{-i\frac{\chi_{b,f}}{2}})C\right) 
    -\arg \left((1+\xi e^{-i\frac{\chi_{b,f}}{2}})A\right)\nonumber \\
    -\arg \left((1+\frac{1}{\xi} e^{i\frac{\chi_{b,f}}{2}})C\right) = \arg \left(\frac{\left(1+\xi e^{i\frac{\chi_{b,f}}{2}}\right)\left(1+\frac{1}{\xi} e^{-i\frac{\chi_{b,f}}{2}}\right)}{\left(1+\xi e^{-i\frac{\chi_{b,f}}{2}}\right)\left(1+\frac{1}{\xi} e^{i\frac{\chi_{b,f}}{2}}\right)}\right) \ .
\end{gather}

Denote $w=\xi e^{i\frac{\chi_{b,f}}{2}}$, hence $\chi=\arg \left(\frac{2+w+\frac{1}{w}}{2+w^*+\frac{1}{w^*}}\right)=2 \arg \left(2+w+\frac{1}{w}\right)$. While $\arg w$ is fixed (up to plus $\pi$), we have control of the value of $\xi$ which can be any real number except zero. Note that $\left(
\frac{\Re\left(w + \frac{1}{w}\right)}{2 \cos\frac{\chi_{b,f}}{2}}
\right)^2
-
\left(
\frac{\Im\left(w + \frac{1}{w}\right)}{2 \sin\frac{\chi_{b,f}}{2}}
\right)^2
= 1$, hence $2+w+\frac{1}{w}$ describes a hyperbola with center at $(2,0)$, and from the origin, we will "see" all the angles between the two asymptotes of the hyperbola in the left half of the plane, i.e. $\arg \left(2+w+\frac{1}{w}\right) \in \left(-\frac{\pi+\chi_{b,f}}{2},\frac{\pi+\chi_{b,f}}{2}\right) $. Hence, after multiplying by $2$ we get that $\chi$ can have any weight value.




To complete our example, there exists a point on the left branch of the hyperbola with argument $\frac{\chi_{b,f}}{2}$ that will yield a $\chi=\chi_{b,f}$ weight. This choice of $A,B,C,D$ will not be a projection onto a 2-qubit maximally entangled state because $AB^*+CD^*=e^{-i\frac{\chi_{b,f}}{2}}\left(\xi |A|^2+\frac{1}{\xi}|C|^2\right)\ne 0$. Yet, this projection will still give us the desired fused weighted graph state (where we keep the original $\chi_{b,f}$ weight) up to single-qubit rotations.

Still in the case $|n(b)|=1$, the set of $A,B,C,D$ such that for graph states, the output state is the desired fused graph states (\ref{eq:General form of A,B,C,D in order to get cluster state after hard transformations}), is not contained in the set of solutions to equations (\ref{eq:first equation for projecting onto weighted graph with non-Bell state projection}),(\ref{eq:second equation for projecting onto weighted graph with non-Bell state projection}) that also satisfy $\chi=\chi_{b,f}$. This hints that a fusion process for graph states that involve projections like (\ref{eq:General form of A,B,C,D in order to get cluster state after hard transformations}) that are projections onto two-qubit maximally entangled states, while retaining its success probability when applied to weighted graph states, will not "really succeed" as the resulting state will not be the desired fused weighted graph state up to single-qubit rotations.

Finally, it is worth parameterizing all the $A,B,C,D$ that give a maximally entangled state (now without any assumption over the size of $n(b)$) --- the condition for that is $M_{ij}'$ (\ref{eq:Mij prime}) being unitary up to multiplication by $\frac{1}{\sqrt{2}}$, meaning:
\begin{gather}
    |A+z^*B|=\sqrt{1-|z|^2}|D| \ , \hspace{0.5cm} |C+z^*D|=\sqrt{1-|z|^2}|B| \ , \hspace{0.5cm}
    AB^*+CD^*+z^*\left(|B|^2+|D|^2\right)=0 \ .
\end{gather}

We can construct such $A,B,C,D$ --- we start with $A',B',C',D'$ such that $\begin{bmatrix}
    A' & B' \\ C' & D'
\end{bmatrix}$ is unitary multiplied by $\frac{1}{\sqrt{2}}$ (this is exactly $M_{ij}'$), and then set:

\begin{gather}
    A= A'- \frac{z^*B'}{\sqrt{1-|z|^2}} \ , \hspace{0.5cm} B=\frac{B'}{\sqrt{1-|z|^2}} \ , \hspace{0.5cm} C=C'-\frac{z^* D'}{\sqrt{1-|z|^2}} \ , \hspace{0.5cm} D=\frac{D'}{\sqrt{1-|z|^2}} \ .
\end{gather}

This is also the full parametrization of $A,B,C,D$ such that the resulting state is maximally entangled.

}

\section{Discussion and Outlook}
\label{Section:Discussion}

In this work, we have extended the analysis of fusion operations to the framework of weighted graph states, which generalize standard graph states by associating continuous phases with their entangling edges. 
Our results have demonstrated that Type-I fusion behaves identically to the unweighted case: it merges two one-dimensional weighted graphs, while preserving both the edge weights, and the success probabilities in the case of unweighted graph states. 
This robustness makes Type-I fusion a reliable operation for constructing large-scale weighted graph states, and supports its use in photonic or other probabilistic architectures. We also showed that the required pool of 2-qubit weighted graph states needed for creating one-dimensional weighted graph states using Type-I fusion, can be generated deterministically by GHZ states or three-qubit weighted graph states, or probabilistically via generalized fusion of Bell pairs. {It would be interesting to try to generalize the creation of 2-qubit weighted graph states from GHZ states --- to a creation of $m$-qubit weighted graph states from $M$-qubit graph states or $M$-qubit GHZ states (or another family of entangled states). This kind of generalization is not straightforward. While two-qubit states are equivalent to each other iff their entanglement entropy is equal, for multi-qubit states the analysis becomes more complicated.}

In contrast, Type-II fusion exhibits fundamental modifications in the weighted setting. 
The creation of the logical qubit necessary for this fusion is restricted to specific weight configurations --- namely, those satisfying either equal edge weights or a fixed $2\pi$ phase relation. 
Moreover, for these configurations, the logical-qubit projection becomes probabilistic, with success rates below one-half. This obstacle can be overcome by creating two adjacent $\pi$-weights in the one-dimensional weighted graph, and applying an $X$ measurement which erases those two weights and leave us with a logical qubit. Hence, this problem only requires us to plan in advance where logical qubits are needed when creating the one-dimensional weighted graph, and the remaining weight configuration is not restricted to equal weights or a $2\pi$ phase relation. 
However, the favorable good-failure feature known from unweighted graph states is lost, since failure outcomes no longer preserve a recoverable graph structure. 
This limitation cannot be overcome by generalizing the fusion process or by modifying the measurement basis: any fusion protocol that retains a nonzero success probability necessarily produces failure outcomes that are not uniform. {That problem can also be overcome by creating in advance many $\pi$-weights, and performing the regular Type-II fusion between qubits that are locally a part of 1-dim regular graph states. This is not ideal, as it requires many additional qubits and Type-I fusions in order to construct these $\pi$-weights. This technique of padding with $\pi$-weights and applying what is locally a Type-II fusion of regular graph states may be relevant for the fusion of more general states than weighted graph states.}

\if{This does not mean that Type-II fusion is necessarily not useful for creating weighted graph state, but rather that  @@}\fi

\if{Beyond these core results, we derived general rules for generating target weighted edges via Y-like measurements, showing that arbitrary edge weights can be engineered deterministically only in special symmetric configurations. }\fi

{To summarize, the loss of good-failure does not obstruct scalable fusion-based creation of weighted graph states; it only imposes an $O\left(n \ln (n)\right)$ qubit overhead, where $n$ is the number of required Type-II fusions needed.}

{\color{black}We also have computed the resulting probabilities and entanglement entropies of generalized fusion applied on weighted graph states, and showed that any fusion process which results in a projecting onto a 2-qubit maximally entangled state in the case of success --- even fusion processes that utilize ancilla qubits --- will retain the success probability when perform on weighted graph states instead of graph states. We remain with the open question, whether there exist a generalized fusion process (without ancilla qubits) with success probability higher than half, using projecting onto 2-qubit non-maximally entangled states.}

We have shown that there are non-Bell projections that will still result in our required fused weighted graph state up to single-qubit rotations. We also showed that in the case of fusing two one-dimensional weighted graph states where one is fused by its edge, a non-Bell projection can create a fused weighted graph state with new weight on the created edge.   

{These findings clarify the operational limits of fusion-based growth of weighted graph states.}
Future work should explore higher-dimensional generalizations, and investigate whether hybrid fusion strategies --- combining Type-I fusions with deterministic edge-weight engineering
\if{, and approaches represented in \cite{yamazaki2025measurementbasedquantumcomputationweighted} (in which a weighted graph state with a uniform weight is made)}\fi
--- can enable scalable preparation of any weighted graph lattices. 
Another promising direction is the study of error-resilient fusion protocols that incorporate feed-forward or adaptive measurements to recover partial entanglement in failure cases. {From an experimental standpoint, one can use this analysis to construct weighted graph states with more than two qubits as those that are realized to date \cite{skosana2026experimentalrealizationphotonicweighted}.
Understanding such mechanisms will be essential for extending the reach of measurement-based quantum computation beyond the unweighted cluster-state paradigm, as well as quantum metrology \cite{skosana2026experimentalrealizationphotonicweighted} and any other field for which weighted graph states may be useful.}

\vspace{0.5cm}

\textbf{Acknowledgments}
We would like to thank Khen Cohen for a valuable discussion.  

\textbf{Funding} - This work is supported in part by the Israeli Science Foundation Excellence Center, the US-Israel Binational Science Foundation, and the Israel Ministry of Science.

\appendix

{\section{Proof that $\ket{\phi_{b_1}}\ket{\phi_{b_2}}$,$e^{-i\phi_{1}\ket{1}_{c_1}\bra{1}_{c_1}}\ket{\phi_{b_1}}\ket{\phi_{b_2}}$,$e^{-i\phi_{2}\ket{1}_{c_2}\bra{1}_{c_2}}\ket{\phi_{b_1}}\ket{\phi_{b_2}}$ and $e^{-i\phi_{1}\ket{1}_{c_1}\bra{1}_{c_1}}e^{-i\phi_{2}\ket{1}_{c_2}\bra{1}_{c_2}}\ket{\phi_{b_1}}\ket{\phi_{b_2}}$ are linear independent}
\label{Appendix:Proof of linear independent}
We follow the footsteps of the proof in \cite{hein2006entanglementgraphstatesapplications} that if $\ket{\phi}$ is a weighted graph state, then all the terms $Z_{a_1}...Z_{a_n}\ket{\phi}$ are linear independent. Notice that $e^{-i\phi_{1}\ket{1}_{c_1}\bra{1}_{c_1}}$ and $e^{-i\phi_{2}\ket{1}_{c_2}\bra{1}_{c_2}}$ commute with all the gates $e^{-i\chi_{a_1 a_2}\ket{1}_{a_1}\ket{1}_{a_2}\bra{1}_{a_1}\bra{1}_{a_2}}$ that are being applied on $\ket{+}_G=\prod_{v\in V} \ket{+}_v$ in order to get the weighted graph state (see eq. (\ref{eq:DefinitionOfWeightedGraphState})), were $G$ is the weighted graph without $a,b,_1,b_2$ and their edges. Hence, it is sufficient to show that $\ket{+}_G$,$e^{-i\phi_{1}\ket{1}_{c_1}\bra{1}_{c_1}}\ket{+}_G$,$e^{-i\phi_{2}\ket{1}_{c_2}\bra{1}_{c_2}}\ket{+}_G$ and $e^{-i\phi_{1}\ket{1}_{c_1}\bra{1}_{c_1}}e^{-i\phi_{2}\ket{1}_{c_2}\bra{1}_{c_2}}\ket{+}_G$ are linear independent, as under applying those unitary gates they will become exactly the states we want to show that are independent. The latter wave functions are:
\begin{gather}
    \frac{1}{2}\left(\ket{0}_{c_1}+\ket{1}_{c_1}\right)\left(\ket{0}_{c_2}+\ket{1}_{c_2}\right) \ , \hspace{0.1cm}
    \frac{1}{2}\left(\ket{0}_{c_1}+e^{-i\phi_1}\ket{1}_{c_1}\right)\left(\ket{0}_{c_2}+\ket{1}_{c_2}\right) \ ,
    \nonumber \\
    \frac{1}{2}\left(\ket{0}_{c_1}+\ket{1}_{c_1}\right)\left(\ket{0}_{c_2}+e^{-i\phi_2}\ket{1}_{c_2}\right) \ , \hspace{0.1cm}
    \frac{1}{2}\left(\ket{0}_{c_1}+e^{-i\phi_1}\ket{1}_{c_1}\right)\left(\ket{0}_{c_2}+e^{-i\phi_2}\ket{1}_{c_2}\right) \ ,
\end{gather}
multiplied by $\ket{+}_{G\setminus \{c_1,c_2\}}$. We simply write those four vectors coordinates (in the basis $\ket{0}_{c_1}\ket{0}_{c_2}$,$\ket{1}_{c_1}\ket{0}_{c_2}$,$\ket{0}_{c_1}\ket{1}_{c_2}$,$\ket{1}_{c_1}\ket{1}_{c_2}$) in a matrix and compute its determinant and show that its different than zero:

\begin{gather}
    \det{\left(\frac{1}{2}\begin{bmatrix}
        1 & 1 & 1 & 1 \\
        1 & e^{-i\phi_1} & 1 & e^{-i\phi_1} \\
        1 & 1 & e^{-i\phi_2} & e^{-i\phi_2} \\
        1 & e^{-i\phi_1} & e^{-i\phi_2} & e^{-i\left(\phi_1+\phi_2\right)}
    \end{bmatrix}\right)}=\frac{e^{-2i\left(\phi_1+\phi_2\right)}\left(e^{i\phi_1}-1\right)^2\left(e^{i\phi_2}-1\right)^2}{16} \ne 0 \ ,
\end{gather}
as desired.
}

\section{Proof of all options for X-like measurement on 3-qubit one-dimensional weighted graph state}
\label{Appendix:ProofOfLogicalQubitWeightedGraphState}
We begin with the state (\ref{eq:WeightedGraphStateAfterApplyngGeneralProjectionOnQubita}) we wish to convert to the state (\ref{eq:DesiredWeightedGraphStateWithb1b2BeingLogicalQubit}) using only 1-qubit rotations on $b_1$ and $b_2$, where there are no other qubits --- the wave functions $\ket{\phi_{b_1}},\ket{\phi_{b_2}}$ are vacuum states and we can erase them from equations (\ref{eq:WeightedGraphStateAfterApplyngGeneralProjectionOnQubita}) and (\ref{eq:DesiredWeightedGraphStateWithb1b2BeingLogicalQubit}). If we write the coefficients in eq. (\ref{eq:WeightedGraphStateAfterApplyngGeneralProjectionOnQubita}) in a matrix:

\begin{gather}
    M=\frac{1}{N_{normalization\hspace{0.1cm}factor}}\begin{bmatrix}
    A+B & A+Be^{-i\chi_1} \\
    A+Be^{-i\chi_2} & A+Be^{-i\left(\chi_1+\chi_2\right)} 
    \end{bmatrix} \ ,
\end{gather}
then our condition is for this matrix to be unitary. Because it is a two-by-two matrix, its (1,1) and (2,2) terms must have the same norm, as its (1,2) and (2,1) terms. This will also be enough to ensure that the two rows have the same norm (and the normalization factor will make this norm one). 

The first equation, $|A+B|=|A+Be^{-i\left(\chi_1+\chi_2\right)}|$ is equivalent to $Re\left(AB^*\right)=Re\left(AB^*e^{i\left(\chi_1+\chi_2\right)}\right)$ (easy to see by squaring both sides and expanding). Because $AB^*$ and $AB^*e^{i\left(\chi_1+\chi_2\right)}$ are from the same norm, they are either the same or complex conjugates of each other, meaning $\arg A - \arg B = \pm \left(\arg A - \arg B +\chi_1+\chi_2\right)$. If $\chi_2\ne -\chi_1$, we must have the minus option, hence $2(\arg B - \arg A )=\chi_1+\chi_2$. If $\chi_2=-\chi_1$ then the original equation holds. 

The second equation, $|A+Be^{-i\chi_1}|=|A+Be^{-i\chi_2}|$ which is equivalent to $Re\left(AB^*e^{i\chi_1}\right)=Re\left(AB^*e^{i\chi_2}\right)$. As before, because $AB^*e^{i\chi_1}$ and $AB^*e^{i\chi_2}$ have the same norm, they are the same or complex conjugate of each other, meaning $\arg A-  \arg B +\chi_1= \pm \left(\arg A - \arg B +\chi_2\right)$. If $\chi_2\ne\chi_1$, we must have the minus option, which is again $2(\arg B - \arg A )=\chi_1+\chi_2$. If $\chi_2=\chi_1$ then the original equation holds.

From those two equations we get that either $2(\arg B - \arg A )=\chi_1+\chi_2$, or $\chi_2=\chi_1$ and $\chi_2=-\chi_1$. The second option means $\chi_1=\chi_2=\pi$ and this is the same case as for the regular graph state.
In the first option, we are now left with the orthogonality condition:

\begin{gather}
    0=\left(A+B\right)\left(A+Be^{-i\chi_1}\right)^*+\left(A+Be^{-i\chi_2}\right)\left(A+Be^{-i\left(\chi_1+\chi_2\right)}\right)^*= \nonumber \\
    =2\left(A+B\right)\left(A+Be^{-i\chi_1}\right)^* \ .
\end{gather}

\if{
\begin{gather}
    0=\left(A+B\right)\left(A+Be^{-i\chi_1}\right)^*+\left(A+Be^{-i\chi_2}\right)\left(A+Be^{-i\left(\chi_1+\chi_2\right)}\right)^*= \nonumber \\
    =\left(\cos{\theta}\pm e^{i\frac{\chi_1+\chi_2}{2}}\sin{\theta}\right)\left(\cos{\theta}\pm e^{i\chi_1}e^{-i\frac{\chi_1+\chi_2}{2}}\sin{\theta}\right)+ \nonumber \\
    +\left(\cos{\theta}\pm e^{-i\chi_2}e^{i\frac{\chi_1+\chi_2}{2}}\sin{\theta}\right)\left(\cos{\theta}\pm e^{i\left(\chi_1+\chi_2\right)}e^{-i\frac{\chi_1+\chi_2}{2}}\sin{\theta}\right)=\nonumber \\
    =2\left(\cos{\theta}\pm e^{i\frac{\chi_1+\chi_2}{2}}\sin{\theta}\right)\left(\cos{\theta}\pm e^{i\frac{\chi_1-\chi_2}{2}}\sin{\theta}\right) \ .
\end{gather}}\fi



Hence, we get that either $A+B=0$ or $A+Be^{-i\chi_1}=0$. If $B=-A$, from $2(\arg B - \arg A )=\chi_1+\chi_2$ we get $\chi_2=-\chi_1$ or there is no solution. If $B=e^{i\chi_1}A$, from $2(\arg B - \arg A )=\chi_1+\chi_2$ we get $\chi_2=\chi_1$ or there is no solution.

\section{Proof of all options for Y-like measurement on 3-qubit one-dimensional weighted graph state}
\label{Appendix:ProofOfCreatingNewEdgeBetweenB1AndB2}
Each of those wave functions, (\ref{eq:WeightedGraphStateAfterApplyngGeneralProjectionOnQubita}) and (\ref{eq:DesiredWeightedGraphStateWithb1b2BeingConnectedByNewEdgeWithWeightPhi}), can be represented by the matrices (now with proper normalization):

\begin{gather}
    M_1=\frac{1}{N_{normalization\hspace{0.1cm}factor}}\begin{bmatrix}
     A+B & A+Be^{-i\chi_1} \\
     A+Be^{-i\chi_2} & A+Be^{-i\left(\chi_1+\chi_2\right)}  
    \end{bmatrix} \ , \hspace{0.2cm} 
    M_2=\frac{1}{2}\begin{bmatrix}
     1 & 1 \\
     1 & e^{-i\phi}  
    \end{bmatrix} \ ,
\end{gather}

where

\begin{gather}
    N_{normalization\hspace{0.1cm}factor}=\sqrt{|A+B|^2+|A+Be^{-i\chi_1}|^2+|A+Be^{-i\chi_2}|^2+|A+Be^{-i\left(\chi_1+\chi_2\right)}|^2} \nonumber \\
    =2\sqrt{1+\frac{1}{2}Re\left(AB^*\left(1+e^{i\chi_1}\right)\left(1+e^{i\chi_2}\right)\right)} \ .
\end{gather}

If we have only the qubits $a,b_1,b_2$ in the original weighted graph state, then single-qubit rotations on $b_1$ and $b_2$ translate to multiplying $M_1$ from left and right by unitary matrices. The question whether by those actions $M_1$ can become $M_2$ is equivalent to $M_1$,$M_2$ having the same singular values. Because they are $2$ by $2$ matrices, this is equivalent to them having the same absolute value of the determinant (as they both have Frobenius norm equal to one). This is equivalent to saying that the entanglement entropy of the resulting state (\ref{eq:WeightedGraphStateAfterApplyngGeneralProjectionOnQubita}) with respect to bipartition into the Hilbert spaces of $b_1$ and $b_2$, determines the weight of the equivalent weighted graph state. The determinants are:

\begin{gather}
    |\det M_2|=\frac{1}{4}|1-e^{-i\phi}| \ , \nonumber \\
    |\det M_1|=\frac{1}{4\left(1+\frac{1}{2}Re\left(AB^*\left(1+e^{i\chi_1}\right)\left(1+e^{i\chi_2}\right)\right)\right)}|A||B||1-e^{-i\chi_1}||1-e^{-i\chi_2}| \ .
\end{gather}

Assuming that the projection onto the state $A\ket{0}_a+B\ket{1}_a$ is the result of measuring a local operator of the qubit $a$, then the other possible projection is onto the state $B^*\ket{0}_a-A^*\ket{1}_a$ up to a global phase. The replacement $A \rightarrow B^* \ , B \rightarrow -A^*$ changes $|\det M_1|$ simply by changing the sign in front of $\frac{1}{2}Re\left(AB^*\left(1+e^{i\chi_1}\right)\left(1+e^{i\chi_2}\right)\right)$ in the denominator. Thus, in order for $|\det M_1|$ to be the same in the two possible outcomes of the measurement (in order to get the same weight in both measurements), we need to impose the condition:

\begin{gather}
    \label{eq:Condition on Re(AB*chi1chi2)}
    Re\left(AB^*\left(1+e^{i\chi_1}\right)\left(1+e^{i\chi_2}\right)\right)=0 \ .
\end{gather}

Notice that:

\begin{gather}
    \arg \left(AB^*\left(1+e^{i\chi_1}\right)\left(1+e^{i\chi_2}\right)\right) = \arg A - \arg B + \frac{\chi_1}{2}+ \frac{\chi_2}{2} \ ,
\end{gather}

where we assume that $-\pi<\chi_1,\chi_2<\pi$. If one of $\chi_1,\chi_2$ equals $\pi$ then $AB^*\left(1+e^{i\chi_1}\right)\left(1+e^{i\chi_2}\right)$ is zero. Thus, condition (\ref{eq:Condition on Re(AB*chi1chi2)}) becomes eq. (\ref{eq:ConditionsOnAandBInOrderToCreateNewWeightEdge}) with the exception when one of $\chi_1,\chi_2$ is $\pi$. Given that this condition holds, the determinant of $M_1$ becomes $|\det M_1|=\frac{1}{4}|A||B||1-e^{-i\chi_1}||1-e^{-i\chi_2}|$. We square both sides of the equation $|\det M_1|=|\det M_2|$ and get:

\begin{gather}
    2\left(1-\cos{\phi}\right)=|A|^2 |B|^2 \cdot 2\left(1-\cos{\chi_1}\right) \cdot 2\left(1-\cos{\chi_2}\right) \ .
\end{gather}

From this we get the value of $\phi$ (\ref{eq:NewWeightBetweenB1AndB2AfterYLikeMeasurement}). By the AM-GM inequality $|A|^2|B|^2\leq\left(\frac{|A|^2+|B|^2}{2}\right)^2=\frac{1}{4}$, with every value of $|A|^2|B|^2$ between $0$ and $\frac{1}{4}$ can be achieved because condition (\ref{eq:ConditionsOnAandBInOrderToCreateNewWeightEdge}) is only on their arguments. Hence, the upper and lower bounds of the value of $\phi$ are given by eq. (\ref{eq:RangeOfTheNewWeightBetweenB1AndB2AfterYLikeMeasurement}).

\if{If we have only the qubits $a,b_1,b_2$ in the original weighted graph state, then single-qubit rotations on $b_1$ and $b_2$ translate to multiplying $M_1$ from left and right by unitary matrices. The question whether by those actions $M_1$ can become $M_2$ is equivalent to $M_1$,$M_2$ having the same singular values. Because there are $2$ by $2$ matrices, this is equivalent to them having the same absolute value of the determinant and the same Frobenius norm. For $M_2$ they are:}\fi

\if{\begin{gather}
    |\det M_2|=\frac{1}{4}|1-e^{-i\phi}| \ , \nonumber \\
    ||M_2||_F=1 \ ,
\end{gather}

and for $M_1$ there are:

\begin{gather}
    |\det M_1|=\frac{1}{4}|A||B||1-e^{-i\chi_1}||1-e^{-i\chi_2}| \ , \nonumber \\
    ||M_1||_F=1+2Re\left(AB^*\left(1+e^{i\chi_1}\right)\left(1+e^{i\chi_2}\right)\right) \ .
\end{gather}

The equality condition of the Frobenius norm yields the condition that $AB^*\left(1+e^{i\chi_1}\right)\left(1+e^{i\chi_2}\right)$ is imaginary. The argument of this number is:

\begin{gather}
    \arg \left(AB^*\left(1+e^{i\chi_1}\right)\left(1+e^{i\chi_2}\right)\right) = \arg A - \arg B + \frac{\chi_1}{2}+ \frac{\chi_2}{2} \ ,
\end{gather}

where we assume that $-\pi<\chi_1,\chi_2<\pi$. If one of them equals $\pi$ then $AB^*\left(1+e^{i\chi_1}\right)\left(1+e^{i\chi_2}\right)$ is zero. Hence we get the condition (\ref{eq:ConditionsOnAandBInOrderToCreateNewWeightEdge}). The other condition can be written as (by taking square):

\begin{gather}
    2\left(1-\cos{\phi}\right)=|A|^2 |B|^2 \cdot 2\left(1-\cos{\chi_1}\right) \cdot 2\left(1-\cos{\chi_2}\right) \ .
\end{gather}

From this we get the value of $\phi$ (\ref{eq:NewWeightBetweenB1AndB2AfterYLikeMeasurement}). In order for such $\phi$ to exist, we must have $0\leq |A|^2 |B|^2 \left(1-\cos{\chi_1}\right) \left(1-\cos{\chi_2}\right) \leq 1$. The left inequality is obviously true, and the right inequality is also always true, because by AM-GM inequality $|A|^2|B|^2\leq\left(\frac{|A|^2+|B|^2}{2}\right)=\frac{1}{4}$ and $\left(1-\cos{\chi_1}\right) \left(1-\cos{\chi_2}\right)\leq2\cdot2=4$. Thus, the only condition is (\ref{eq:ConditionsOnAandBInOrderToCreateNewWeightEdge}), and we can deduce the upper and lower bounds of the value of $\phi$ (\ref{eq:NewWeightBetweenB1AndB2AfterYLikeMeasurement}).}\fi

{\section{Erasing added $\pi$-weights chain}
\label{appendix:Erasing added pi weights chain}
Assume we have succeeded in fusing the two padded 1-dim weighted graph states. Now, our aim is to erase all the remaining $\pi$-weights as in figure \ref{fig:Erasingpi}.a. We start by applying $Y$ measurements on the neighbors of the center qubit $e$, which each time will erase those qubits and will connect $e$ to the next qubits in the four chains, until we have two $\pi$-weights in each direction as in figure \ref{fig:Erasingpi}.b. Then, we will measure the neighbors of $e$, $x_1,x_2,x_3,x_4$, in $X$ basis, which will erase them, and the next qubits in line, $y_1,y_2,y_3,y_4$, will form a logical qubit together with $e$, meaning $\ket{0}_L=\ket{0}_e\prod_{i=1}^{i=4}\ket{0}_i$ and $\ket{1}_L=\ket{1}_e\prod_{i=1}^{i=4}\ket{1}_i$, as displayed in figure \ref{fig:Erasingpi}.c. Now, by eq. (\ref{eq:RecursiveDefinitionOfWeightedGraphState}), the wave function can be written as:

\begin{gather}
    \ket{\psi}=\frac{1}{\sqrt{2}}\left(\ket{0}_L\ket{\phi}+\ket{1}_L\prod_{l\in n(L)}e^{-i\chi_{l,n(L)}\ket{1}_l\bra{1}_l}\ket{\phi}\right) \ .
    \label{eq:erasing pi weights, final fight}
\end{gather}

When we now apply $X$ measurement on $y_1$, we apply on eq. (\ref{eq:erasing pi weights, final fight}) one of the projections (up to normalization) $\bra{0}_{y_1}\pm \bra{1}_{y_1}$, which will just erase $y_1$ from $L$ up to applying $Z_e$. Doing the same with $y_2,y_3,y_4$ will result in the desired fused weighted graph state as displayed in figure \ref{fig:Erasingpi}.d, because in eq. (\ref{eq:erasing pi weights, final fight}), $\ket{0}_L,\ket{1}_L$ will just be replaced by $\ket{0}_e,\ket{1}_e$.


\begin{figure}
    \centering
    \includegraphics[width=1\linewidth]{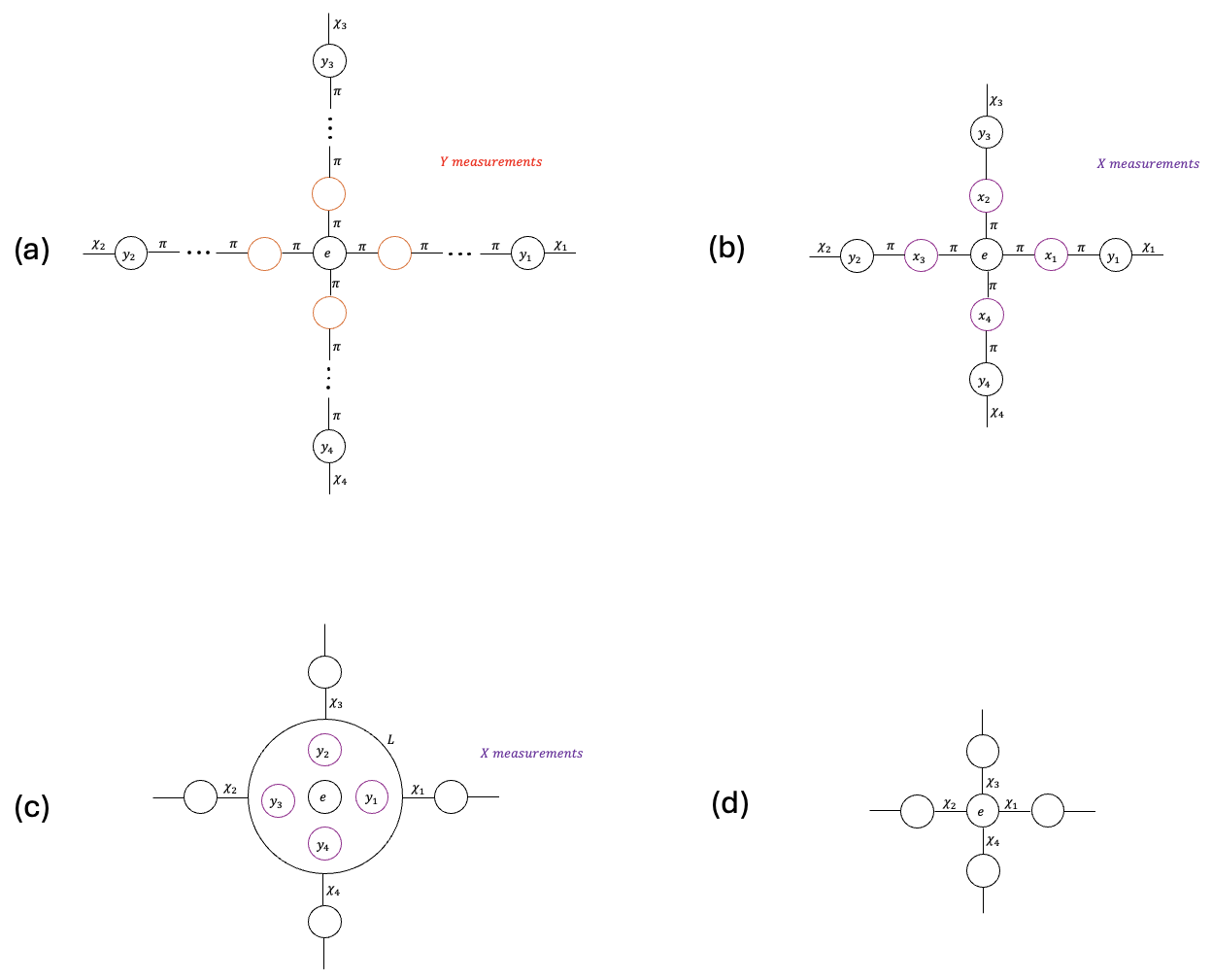}
    \caption{{Erasing all the additional $\pi$-weights. We start with the state (a). By Y measurements on the neighbors of the center qubit we erase $\pi$-weights and qubits until we remain with two $\pi$-weights in each direction (b). Now, we apply $X$ measurements on the neighbors of the center qubit, erasing all the $\pi$-weights and creating a logical qubit instead of the center qubit (c). Finally, we apply $X$ measurements on all the qubits that make up the logical qubit but one, and we end up with the desired fused weighted graph state (d).}}
    \label{fig:Erasingpi}
\end{figure}


}

\section{Proof of conditions (\ref{eq:first equation for projecting onto weighted graph with non-Bell state projection}) and (\ref{eq:second equation for projecting onto weighted graph with non-Bell state projection})}
\label{appendix:Proof of creating weighted graph state by non-Bell projection}
In the basis $\ket{0}_e\ket{0}_f,\ket{0}_e\ket{1}_f,\ket{1}_e\ket{0}_f,\ket{1}_e\ket{1}_f$ the operator $T_{e,f}$ (\ref{eq:The 2-qubits gate between e and n(b) for graph state}) takes the form (up to normalization):

\begin{gather}
    \begin{bmatrix}
        A+B & 0 & 0 & 0 \\
        0 & A+Be^{-i\chi_{b,f}} & 0 & 0 \\
        0 & 0 & C+D & 0 \\
        0 & 0 & 0 & C+De^{-i\chi_{b,f}}    
    \end{bmatrix} \ ,
    \label{eq:State ABCD in basis 00 01 10 11}
\end{gather}

which means that $T_{e,f}$ being unitary is equivalent to:

\begin{gather}
    |A+B|=|A+Be^{-i\chi_{b,f}}|=|C+D|=|C+De^{-i\chi_{b,f}}| \ .
\end{gather}

Taking square on $|A+B|=|A+Be^{-i\chi_{b,f}}|$ which is analogous to the equation $|A+B|=|A+Be^{-i\left(\chi_1+\chi_2\right)}|$ in appendix \ref{Appendix:ProofOfLogicalQubitWeightedGraphState}, we have $\arg A -\arg B = \pm \left(\arg A - \arg B + \chi_{b,f}\right)$. Because $\chi_{b,f}\ne0$ we get $\arg B - \arg A = \frac{\chi_{b,f}}{2}+0\hspace{0.1cm}or\hspace{0.1cm}\pi$, and by applying the same argument on $C,D$ we get the same $\arg D - \arg C = \frac{\chi_{b,f}}{2}+0\hspace{0.1cm}or\hspace{0.1cm}\pi$.

\if{\begin{gather}
    \arg B - \arg A = \arg D - \arg C = \frac{\chi_{b,f}}{2}+0\hspace{0.1cm}or\hspace{0.1cm}\pi \ ,
\end{gather}}\fi

Now we are left with the equation $|A+B|=|C+D|$. After squaring both sides, we get $|A|^2+|B|^2+2|A||B|\cos{\left(\arg B - \arg A\right)}=|C|^2+|D|^2+2|C||D|\cos{\left(\arg D - \arg C\right)}$. Substituting $\arg B - \arg A$ and $\arg D - \arg C$ we get $|A|^2+|B|^2\pm 2|A||B|\cos{\frac{\chi_{b,f}}{2}}=|C|^2+|D|^2\pm 2|C||D|\cos{\frac{\chi_{b,f}}{2}}$ where the signs in $\pm$ are not necessarily the same on both sides, but determined by the values of the argument differences $\arg B - \arg A$ and $\arg D - \arg C$; if $\arg B - \arg A=\frac{\chi_{b,f}}{2}$ then the sign before $2|A||B|\cos{\frac{\chi_{b,f}}{2}}$ is plus, else the sign is minus, and the same holds for the sign before $2|C||D|\cos{\frac{\chi_{b,f}}{2}}$.

\if{One can also compute the resulting weight between $e$ and $f$, which is (see proof A3 in proof appendix of \cite{rimock2024generalizedtypeiifusion}):

\begin{gather}
    \chi=\arg \left(A+B\right) + \arg \left(C+De^{-i\chi_{b,f}}\right)-\arg \left(A+Be^{-i\chi_{b,f}}\right)-\arg \left(C+D\right)
\end{gather}}\fi

\clearpage

\bibliography{ref}

\end{document}